\documentclass[final,5p,twocolumn]{elsarticle}

\usepackage{lscape}

\usepackage{graphics}

\usepackage{amssymb}

\usepackage{color}
\usepackage{fixltx2e}

\newcommand{\captionfonts}{\normalsize}
\makeatletter  
\long\def\@makecaption#1#2{%
  \vskip\abovecaptionskip
  \sbox\@tempboxa{{\captionfonts #1: #2}}%
  \ifdim \wd\@tempboxa >\hsize
    {\captionfonts #1: #2\par}
  \else
    \hbox to\hsize{\hfil\box\@tempboxa\hfil}%
  \fi
  \vskip\belowcaptionskip}
\makeatother   

\usepackage{natbib}
\bibpunct{(}{)}{;}{a}{}{,}

\journal{Astroparticle Physics}

\begin{document}





\vspace{-10cm}

\title{The major upgrade of the MAGIC telescopes, Part I: The hardware
improvements and the commissioning of the system}



%
\author[a]{J.~Aleksi\'c}
\author[b]{S.~Ansoldi}
\author[c]{L.~A.~Antonelli}
\author[d]{P.~Antoranz}
\author[e]{A.~Babic}
\author[f]{P.~Bangale}
\author[a]{M.~Barcel\'o}
\author[g]{J.~A.~Barrio}
\author[h,*]{J.~Becerra Gonz\'alez}
\author[i]{W.~Bednarek}
\author[j]{E.~Bernardini}
\author[b]{B.~Biasuzzi}
\author[k]{A.~Biland}
\author[y]{M.~Bitossi}
\author[a]{O.~Blanch}
\author[g]{S.~Bonnefoy}
\author[c]{G.~Bonnoli}
\author[f]{F.~Borracci}
\author[l,**]{T.~Bretz}
\author[m]{E.~Carmona}
\author[c]{A.~Carosi}
\author[z]{R.~Cecchi}
\author[f]{P.~Colin}
\author[h]{E.~Colombo}
\author[g]{J.~L.~Contreras}
\author[o]{D.~Corti}
\author[a]{J.~Cortina}
\author[c]{S.~Covino}
\author[d]{P.~Da Vela}
\author[f]{F.~Dazzi}
\author[b]{A.~De Angelis}
\author[j]{G.~De Caneva}
\author[b]{B.~De Lotto}
\author[n]{E.~de O\~na Wilhelmi}
\author[m]{C.~Delgado Mendez}
\author[f]{A.~Dettlaff}
\author[e]{D.~Dominis Prester}
\author[l]{D.~Dorner}
\author[o]{M.~Doro}
\author[p]{S.~Einecke}
\author[l]{D.~Eisenacher}
\author[l]{D.~Elsaesser}
\author[g]{D.~Fidalgo}
\author[f]{D.~Fink}
\author[g]{M.~V.~Fonseca}
\author[q]{L.~Font}
\author[p]{K.~Frantzen}
\author[f]{C.~Fruck}
\author[r]{D.~Galindo}
\author[h]{R.~J.~Garc\'ia L\'opez}
\author[j]{M.~Garczarczyk}
\author[q]{D.~Garrido Terrats}
\author[q]{M.~Gaug}
\author[a,j]{G.~Giavitto}
\author[e]{N.~Godinovi\'c}
\author[a]{A.~Gonz\'alez Mu\~noz}
\author[j]{S.~R.~Gozzini}
\author[f]{W.~Haberer}
\author[n,***]{D.~Hadasch}
\author[s]{Y.~Hanabata}
\author[s]{M.~Hayashida}
\author[h]{J.~Herrera}
\author[k]{D.~Hildebrand}
\author[f]{J.~Hose}
\author[e]{D.~Hrupec}
\author[i]{W.~Idec}
\author[a]{J.~M.~Illa}
\author[t]{V.~Kadenius}
\author[f]{H.~Kellermann}
\author[k]{M.~L.~Knoetig}
\author[s]{K.~Kodani}
\author[s]{Y.~Konno}
\author[f]{J.~Krause}
\author[s]{H.~Kubo}
\author[s]{J.~Kushida}
\author[c]{A.~La Barbera}
\author[e]{D.~Lelas}
\author[g]{J.~L.~Lemus}
\author[l]{N.~Lewandowska}
\author[t,&]{E.~Lindfors}
\author[c]{S.~Lombardi}
\author[b]{F.~Longo}
\author[g]{M.~L\'opez}
\author[a]{R.~L\'opez-Coto}
\author[a]{A.~L\'opez-Oramas}
\author[g]{A.~Lorca}
\author[f,****]{E.~Lorenz}
\author[g]{I.~Lozano}
\author[u]{M.~Makariev}
\author[j]{K.~Mallot}
\author[u]{G.~Maneva}
\author[b,af]{N.~Mankuzhiyil}
\author[l]{K.~Mannheim}
\author[c]{L.~Maraschi}
\author[r]{B.~Marcote}
\author[o]{M.~Mariotti}
\author[a]{M.~Mart\'inez}
\author[f,s]{D.~Mazin\corref{cor1}}
\ead{mazin@mpp.mpg.de}
\cortext[cor1]{Corresponding author}

\author[f]{U.~Menzel}
\author[d]{J.~M.~Miranda}
\author[f]{R.~Mirzoyan}
\author[a]{A.~Moralejo}
\author[r]{P.~Munar-Adrover}
\author[s]{D.~Nakajima}
\author[o]{M.~Negrello}
\author[t]{V.~Neustroev}
\author[i]{A.~Niedzwiecki}
\author[t,&]{K.~Nilsson}
\author[s]{K.~Nishijima}
\author[f]{K.~Noda}
\author[s]{R.~Orito}
\author[p]{A.~Overkemping}
\author[o]{S.~Paiano}
\author[b]{M.~Palatiello}
\author[f]{D.~Paneque}
\author[d]{R.~Paoletti}
\author[r]{J.~M.~Paredes}
\author[r]{X.~Paredes-Fortuny}
\author[b,&&]{M.~Persic}
\author[t]{J.~Poutanen}
\author[v]{P.~G.~Prada Moroni}
\author[k,&&&]{E.~Prandini}
\author[e]{I.~Puljak}
\author[t]{R.~Reinthal}
\author[p]{W.~Rhode}
\author[r]{M.~Rib\'o}
\author[a]{J.~Rico}
\author[f]{J.~Rodriguez Garcia}
\author[l]{S.~R\"ugamer}
\author[s]{T.~Saito}
\author[s]{K.~Saito}
\author[g]{K.~Satalecka}
\author[o]{V.~Scalzotto}
\author[g]{V.~Scapin}
\author[o]{C.~Schultz}
\author[f]{J.~Schlammer}
\author[f]{S.~Schmidl}
\author[f]{T.~Schweizer}
\author[t]{A.~Sillanp\"a\"a}
\author[a]{J.~Sitarek}
\author[e]{I.~Snidaric}
\author[i]{D.~Sobczynska}
\author[l]{F.~Spanier}
\author[c]{A.~Stamerra}
\author[l]{T.~Steinbring}
\author[l]{J.~Storz}
\author[f]{M.~Strzys}
\author[t]{L.~Takalo}
\author[s]{H.~Takami}
\author[c]{F.~Tavecchio}
\author[g]{L.~A.~Tejedor}
\author[u]{P.~Temnikov}
\author[e]{T.~Terzi\'c}
\author[h]{D.~Tescaro\corref{cor1}}
\ead{diego.tescaro@gmail.com}

\author[f]{M.~Teshima}
\author[p]{J.~Thaele}
\author[l]{O.~Tibolla}
\author[w]{D.~F.~Torres}
\author[f]{T.~Toyama}
\author[x]{A.~Treves}
\author[k]{P.~Vogler}
\author[f]{H.~Wetteskind}
\author[h]{M.~Will}
\author[r]{R.~Zanin}

\address[a]{IFAE, Campus UAB, E-08193 Bellaterra, Spain}
\address[b]{Universit\`a di Udine, and INFN Trieste, I-33100 Udine, Italy}
\address[c]{INAF National Institute for Astrophysics, I-00136 Rome, Italy}
\address[d]{Universit\`a  di Siena, and INFN Pisa, I-53100 Siena, Italy}
\address[e]{Croatian MAGIC Consortium, Rudjer Boskovic Institute, University of Rijeka and University of Split, HR-10000 Zagreb, Croatia}
\address[f]{Max-Planck-Institut f\"ur Physik, D-80805 M\"unchen, Germany}
\address[g]{Universidad Complutense, E-28040 Madrid, Spain}
\address[h]{Inst. de Astrof\'isica de Canarias, E-38200 La Laguna, Tenerife, Spain}
\address[i]{University of \L\'od\'z, PL-90236 Lodz, Poland}
\address[j]{Deutsches Elektronen-Synchrotron (DESY), D-15738 Zeuthen, Germany}
\address[k]{ETH Zurich, CH-8093 Zurich, Switzerland}
\address[l]{Universit\"at W\"urzburg, D-97074 W\"urzburg, Germany}
\address[m]{Centro de Investigaciones Energ\'eticas, Medioambientales y Tecnol\'ogicas, E-28040 Madrid, Spain}
\address[n]{Institute of Space Sciences, E-08193 Barcelona, Spain}
\address[o]{Universit\`a di Padova and INFN, I-35131 Padova, Italy}
\address[p]{Technische Universit\"at Dortmund, D-44221 Dortmund, Germany}
\address[q]{Unitat de F\'isica de les Radiacions, Departament de F\'isica, and CERES-IEEC, Universitat Aut\`onoma de Barcelona, E-08193 Bellaterra, Spain}
\address[r]{Universitat de Barcelona, ICC, IEEC-UB, E-08028 Barcelona, Spain}
\address[s]{Japanese MAGIC Consortium, KEK, Department of Physics and Hakubi Center, Kyoto University, Tokai University, The University of Tokushima, ICRR, The University of Tokyo, Japan}
\address[t]{Finnish MAGIC Consortium, Tuorla Observatory, University of Turku and Department of Physics, University of Oulu, Finland}
\address[u]{Inst. for Nucl. Research and Nucl. Energy, BG-1784 Sofia, Bulgaria}
\address[v]{Universit\`a di Pisa, and INFN Pisa, I-56126 Pisa, Italy}
\address[w]{ICREA and Institute of Space Sciences, E-08193 Barcelona, Spain}
\address[x]{Universit\`a dell'Insubria and INFN Milano Bicocca, Como, I-22100 Como, Italy}
\address[y]{European Gravitational Observatory, I-56021 S. Stefano a Macerata, Italy}
\address[z]{Universit\`a di Siena and INFN Siena, I-53100 Siena, Italy}
\address[*]{now at NASA Goddard Space Flight Center, Greenbelt, MD 20771, USA and Department of Physics and Department of Astronomy, University of Maryland, College Park, MD 20742, USA}
\address[**]{now at Ecole polytechnique f\'ed\'erale de Lausanne (EPFL), Lausanne, Switzerland}
\address[***]{now at Institut f\"ur Astro- und Teilchenphysik, Leopold-Franzens- Universit\"at Innsbruck, A-6020 Innsbruck, Austria}
\address[&]{now at Finnish Centre for Astronomy with ESO (FINCA), Turku, Finland}
\address[af]{now at Astrophysics Science Division, Bhabha Atomic Research Centre, Mumbai 400085, India}
\address[&&]{also at INAF-Trieste}
\address[&&&]{also at ISDC - Science Data Center for Astrophysics, 1290, Versoix (Geneva)}
\address[****]{deceased}

\vspace{-2cm}

\begin{abstract}

\end{abstract}




\maketitle

\section*{Abstract}
The MAGIC telescopes are two Imaging Atmospheric Cherenkov Telescopes (IACTs)
located on the Canary island of La Palma. The telescopes are designed to measure
Cherenkov light from air showers initiated by gamma rays in the energy regime 
from around 50\,GeV to more than 50\,TeV. 
The two telescopes were built in 2004 and 2009, respectively, with different cameras, triggers
and readout systems. 
In the years 2011-2012 the MAGIC collaboration undertook a major upgrade to
make the stereoscopic system uniform, improving its overall performance and
easing its maintenance. In particular, the camera, the receivers and the trigger of the first telescope 
were replaced and the readout of the two telescopes was upgraded. 
This paper (Part~I) describes the 
details of the upgrade as well as the basic performance parameters of MAGIC such as raw data treatment, 
linearity in the electronic chain and sources of noise.
In Part~II, we describe the physics performance of the upgraded system.

\section{Introduction}
\label{sec:intro}

\begin{figure}
    \includegraphics[width=0.49\textwidth]{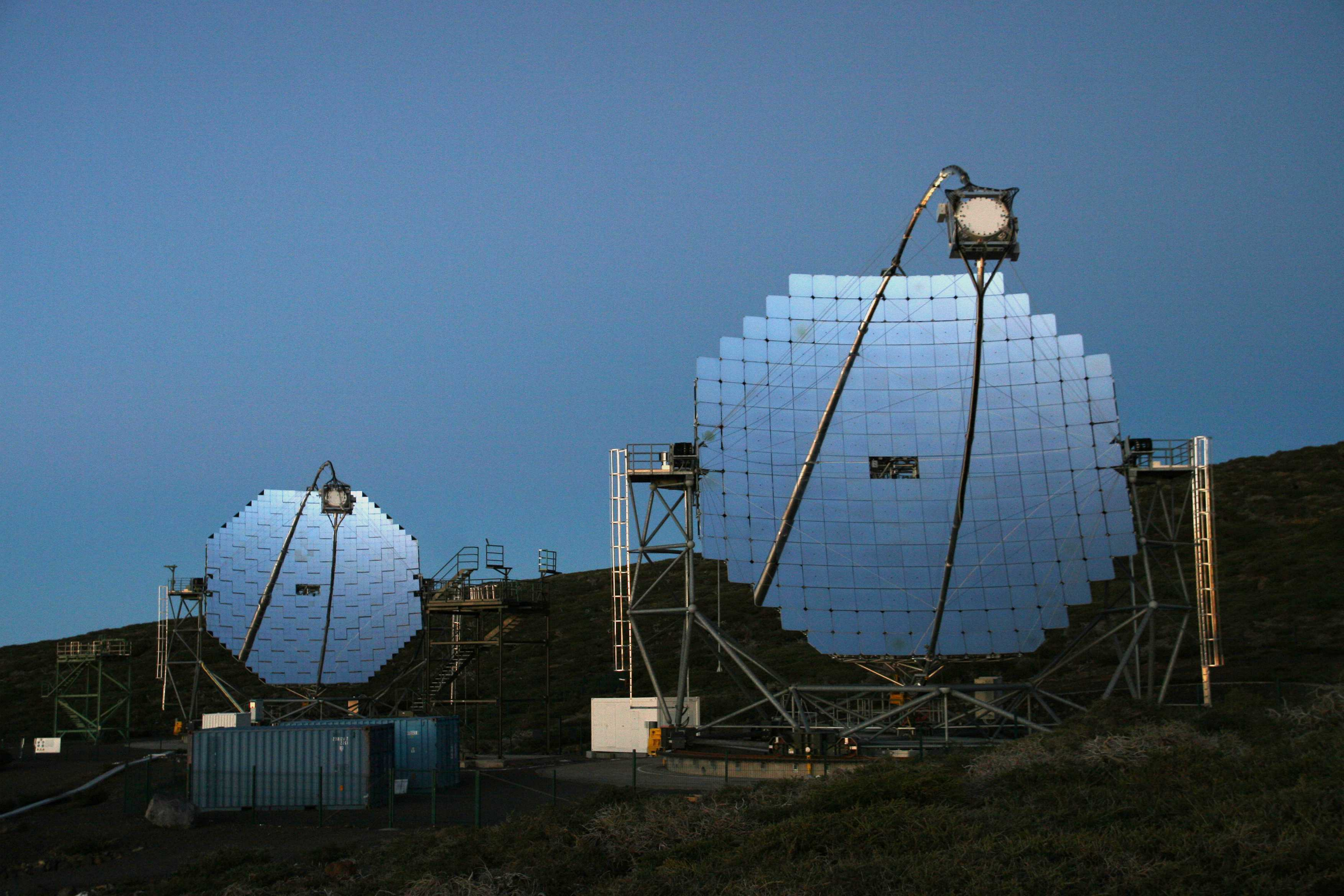}
    \caption{The two 17~m diameter MAGIC telescope system operating at the
Roque de los Muchachos observatory in La Palma.
The telescope in front is MAGIC-II.}
    \label{2-Magics-photo}
\end{figure}
\setcounter{footnote}{0}   
 
MAGIC (see Fig.~\ref{2-Magics-photo}) is a stereoscopic system of two Imaging 
Atmospheric Cherenkov Telescopes (IACTs) located at 2200~m a.s.l. in the observatory
of Roque de los Muchachos in La Palma, Canary Islands (Spain). 
Together with the H.E.S.S.\ IACTs in Namibia \citep{aharonian:2006:hess:crab}
and the VERITAS IACTs in Arizona \citep{holder:2008a}, MAGIC 
is the most sensitive instrument for high-energy gamma-ray astrophysics
in the range between few tens of GeVs and tens of TeVs.

Contrary to {\bf optical} telescopes, IACTs observe dim ($\sim$100 photons / m$^2$ / TeV) short ($\sim$\,ns) flashes produced by extended air showers developing in the atmosphere \citep[see reviews by, e.g.,][]{2009IACT_jhinton,lorenz:2012a}.
The light, mostly emitted in the UV and optical wave bands, is produced via Cherenkov 
radiation from the charged particles of the atmospheric shower, which travel faster than the light in the air.
The amount of Cherenkov light and its angular and spatial distribution
carry information about the energy and incoming direction of the primary cosmic rays and $\gamma$ rays, 
which is reconstructed analyzing the image formed on the focal plane of the IACTs.
The images roughly resemble an ellipse, 
whose brightness, geometrical size, and orientation represent the most basic parameters used in the subsequent data analysis \citep[see][for details]{hillas:1985:images}.
The telescopes are self-triggered by multiple (neighbor) pixels above a certain signal threshold. 
Because the Cherenkov light flashes from air showers are very short, typically few nanoseconds long, 
the use of extremely fast and sensitive light sensors, typically photomultiplier tubes (PMTs), and fast electronics for the trigger and signal sampling is the key to discriminate
the shower light from fluctuations of the night sky background.
The amount of Cherenkov photons reaching the pixels is reconstructed from the signal charge in the PMTs, by analyzing the ultra-fast sampled snapshot of the signal pulse.
An ``extraction" method, that basically sums the ADC counts in a certain time (sliding) window, 
provides a rough signal charge {\bf per} channel, which, after a calibration procedure, is converted into the number of photons at the camera plane \citep{zanin:2013:magic:mars}.
A coincidence (stereo) trigger among individual telescopes minimizes spurious events triggered by the night sky background light, triggers by the so-called afterpulsing effect of the PMTs or by single local muons flashing only one telescope. 
Moreover, in the so-called stereoscopic reconstruction,
{\bf multiple images of the air shower allow the energy and the incoming direction of the primary 
$\gamma$ ray to be more precisely reconstructed.}

The two MAGIC telescopes started operation 5 years apart (MAGIC-I in 2004 and MAGIC-II in 2009, respectively), and the second telescope was an ``improved clone'' of the first one.
The main reasons for differences were funding constraints during the building of the first telescope and the technological progress that took place in the years between the design of the two telescopes. 
The major goal of the telescopes is a lowest possible energy threshold, which is achieved
through fine pixelated cameras, fast sampling electronics and a large mirror area.
The second goal is a fast repositioning speed in order to catch rapid transient events such as Gamma-Ray Bursts,
which is achieved through a light weight ($<$70\,tons) telescope structure made out of 
reinforced carbon fibre tubes. The structure requires an automatic mirror control (AMC)
to maintain the best possible optical point spread function at different zenith angles of observations
\citep{lorenz:2004a,doro:2011a}.
The readout {\bf and the trigger electronics} are located in a dedicated counting house, where the signals transmitted via optical fibers from the camera{\bf s} are received.
A {\bf difference} in transit time between signals in different channels
(mainly due to different high voltages applied to PMTs) is corrected online at
trigger level {\bf by means of adjustable delay lines} to minimize the
needed trigger gate and offline for the reconstruction of the signal arrival
time and charge.  The achieved energy threshold is as low as $\sim$50\,GeV at
the trigger level for observations at zenith angles below 25$^{\circ}$
\citep[see Fig.~6 in][]{aleksic:2014:magic:upgradePartII}.  This energy
threshold is achieved by means of {\bf a} digital trigger. Using the so-called
sum-trigger, {\bf it} is possible to reach an even lower energy threshold
\citep{aliu:2008:crabpulsar}, and a new version of the sum-trigger is currently
under commissioning \citep{jezabel:sumtrigger:icrc:2013a}.  The repositioning
speed is maintained throughout the years to be $\sim$\,25\,s for a
180$^{\circ}$ rotation in azimuth.

While the above mentioned concepts made the two MAGIC telescopes look very similar there were few 
important design differences between MAGIC-I and II before the upgrade described in this paper.
{\bf Funding permitted to equip the entire MAGIC-II field of view (FoV) homogeneously with
small 1 inch PMTs, compared to the mixed 1 and 2 inch pixel configuration of
the MAGIC-I camera. The active trigger area, which in all MAGIC cameras is
limited to a central area in the FoV, was enlarged from $\sim$\,0.9\,$\deg$ radius (trigger
area of the old MAGIC-I camera) to $\sim\,$1.2\,$\deg$ radius (in the MAGIC-II camera), still using the same
trigger electronics as for the \mbox{MAGIC-I} camera but reducing the size of
overlapping sectors (see Section~\ref{sec:trigger}). The main motivation for
enlarging the sensitive trigger area was to adapt to the stereo approach and
increase sensitivity to extended $\gamma$-ray sources as well as a more
suitable usage of the so-called wobble mode \citep[pointing to a source of
interest at some $0.4\,\deg$ off-center,][]{fomin:1994:wobble} for a better
background estimation.}

In detail, the main resulting differences between the two telescopes were the following ones:
\begin{itemize}
   \item The camera of the MAGIC-I telescope consisted of 577 PMTs divided in 397 small PMTs, 1 inch diameter each, in the inner part of the camera and 180 large PMTs, 2 inch diameter each, in the outer part.
The FoV the camera was 3.5$^\circ$.
The camera of MAGIC-II consists of 1039 PMTs, all 1 inch diameter, and has the same FoV as the first camera.
   \item The region of the MAGIC-II camera exploited {\bf for} the trigger was 1.7 times larger than the one of MAGIC-I. 
   \item {\bf The} MAGIC-I readout was based on an optical multiplexer and off-the-shelf Flash Analog to Digital Converters (FADCs) \citep[MUX-FADC,][]{bartko:2005a}, which was robust and had an excellent performance but was expensive and bulky. 
The readout of MAGIC-II was based on the DRS2 chip\footnote{See http://drs.web.psi.ch/.} (compact and inexpensive but performing quite worse in terms of intrinsic noise, dead time and linearity compared to the MUX-FADC system).
   \item The receiver boards of MAGIC-I (see Sec.~\ref{sec:receiverboards}), the part of the electronics responsible to convert the optical signals coming from the camera and to generate the level zero trigger signal, lacked programmability. 
They were also showing high failure rate, mainly due to aging. 
\end{itemize} 

In 2011-2012 MAGIC underwent a major upgrade
program to improve and to unify the stereoscopic system of the two telescopes.
Most importantly, the camera of MAGIC-I was replaced by a new one,
the readout of the two telescopes replaced by a more modern system,
and the trigger area of the MAGIC-I {\bf camera} was increased to match the one of MAGIC-II.
Table~\ref{tab:specs} provides a brief summary of the most relevant hardware characteristics of the telescopes before and after the upgrade.
This paper (Part~I) describes the motivation for the upgrade, its main steps, 
the commissioning of the system and the low level performance of MAGIC. 
In Part~II \citep{aleksic:2014:magic:upgradePartII} 
we describe the physics performance of the upgraded system.

\begin{table}[htp]
\begin{center}
\begin{tabular}{l|cc|c}
{} &  \multicolumn{2}{c|}{Before Upgrade} & After Upgrade\\
Parameter & M-I & M-II & M-I/M-II \\
\hline
Digitizer type & Aquiris$^\dagger$ & DRS2 & DRS4 \\
ADC res. (bits) & 10 & 12 & 14 \\
Sampling (GS/s) & 2.00 & 2.05 & 2.05  \\
Dead time ($\mu$s) & 25 & 500 & 27 \\
Camera shape & hexagonal & round & round \\
Total pixels  & 577($180^\ddagger$) & 1039 & 1039 \\
N trigger pixels  & 325 & 547 & 547 \\
Trig. area ($\deg^2$) & 2.55 & 4.30 & 4.30 \\
{\bf Field of View ($\deg$)} & {\bf 3.5} & {\bf 3.5} & {\bf 3.5} \\
\end{tabular}
\end{center}
\caption{Hardware specifications of the MAGIC system before and after the upgrade (``M-" stands for MAGIC-). $\dagger$: Commercial FADC, multiplexed. $\ddagger$:~Number of outer large (2 inch) pixels.}
\label{tab:specs}
\end{table}%

\section{Motivation for the upgrade}

There were three main motivations for the upgrade of the MAGIC system.
The first one was the wish to improve the stereoscopic performance of the MAGIC system.
Several key parameters were targeted for improvement:
\begin{itemize}
 \item {\textbf {The low energy performance.}}
The performance of MAGIC to the lowest accessible energies was limited by the electronic noise
in the DRS2 system of the MAGIC-II telescope. With a lower noise system the analysis energy threshold can be lowered, and the performance close to the threshold can be improved.
 \item {\textbf {The flux sensitivity to extended sources.}}
The small trigger area of the MAGIC-I telescope (1 degree diameter) was hindering a study of extended Galactic gamma-ray sources, with angular sizes $\geq0.3^{\circ}$.
A 70\% larger trigger region, the same as in the MAGIC-II telescope, allows to measure an extended source up to $\sim\,0.5^{\circ}$ extension, and a better control of the background region. 
 \item {\textbf {The dead time of the system.}}
Due to the intrinsic constraints of the DRS2 based readout of MAGIC-II, 
the dead time of the system was 500\,$\mu$s for every recorded event, which was translating into a $\sim12$\% dead time. 
Reducing the dead time per event by a factor of $\sim$10 was one of the goals of the upgrade 
in order to effectively gain $\sim12$\% of observation time.
 \item {\textbf {The angular resolution for gamma rays.}} 
Replacing the MAGIC-I camera with one containing small pixels only, the image
parameters can be better determined, which helps in the reconstruction of the
primary $\gamma$-ray characteristics such as their incoming direction.
\end{itemize}

The second main motivation was a reduction of any downtime due to technical problems.
This was achieved by upgrading the {\bf subsystems} more prone to failure and implementing many diagnostic and online monitoring tools to immediately alert the shifters and subsystem experts in case of any malfunctioning.
{\bf S}pecial attention was given to producing and storing in La Palma a sufficient amount of spares for most of the hardware.
 
The third motivation was to reduce the manpower and expertise needed to run MAGIC in the following years. 
Less diversification of the {\bf subsystems} reduces the typologies of problems that the shifters may encounter during operation, and also reduces the overhead for eventual troubleshooting from the experts.

\section{Individual parts of the upgrade}
\label{sec:trigger}

\begin{figure}[phtb]
\begin{center}
\includegraphics[width=\linewidth]{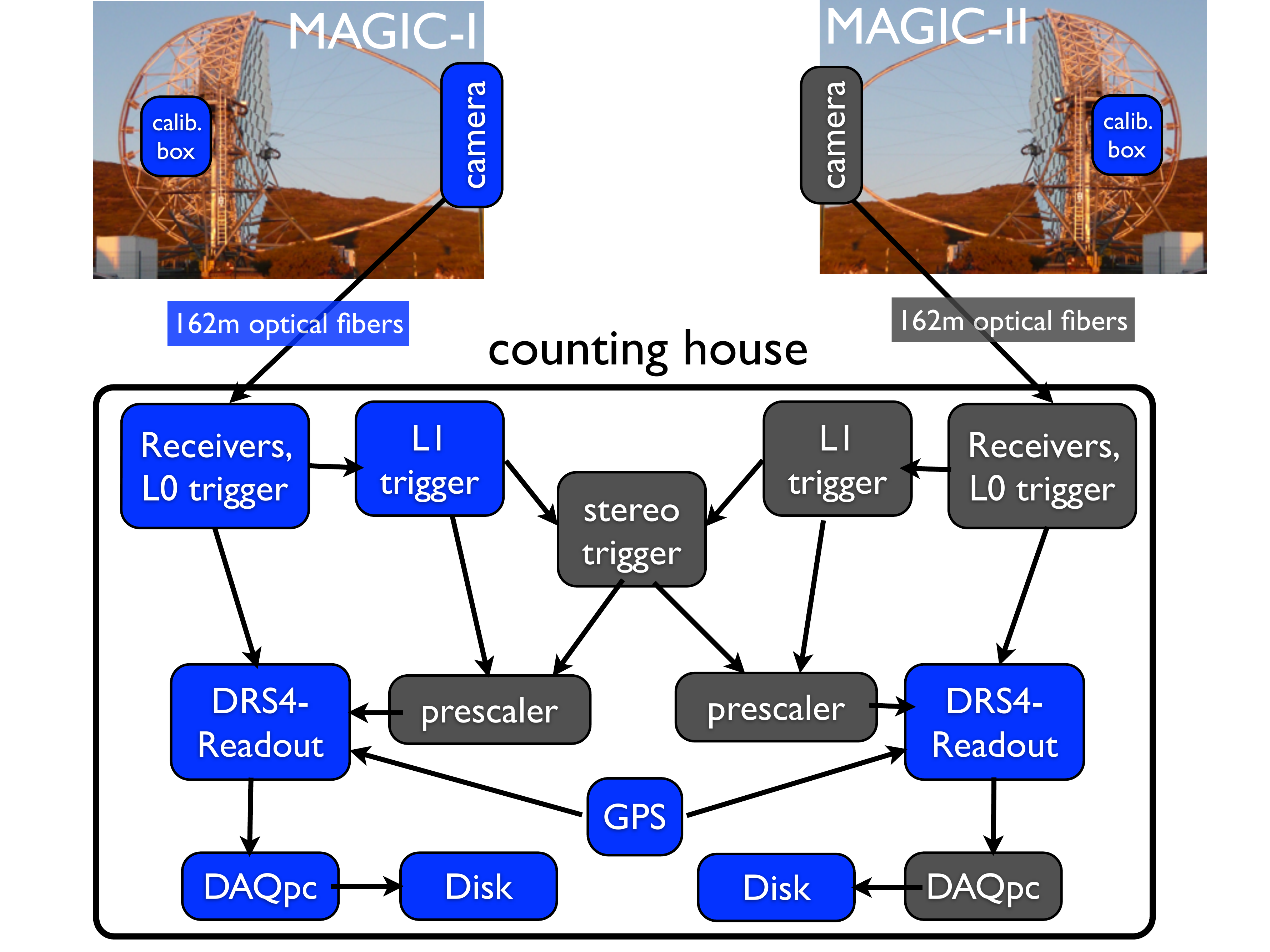}
\caption{
Schematic view of the readout and trigger chain of the MAGIC telescopes.
The blocks in the blue boxes have been replaced and commissioned during the upgrade.}
\label{fig:MAGIC-Scheme}
\end{center}
\end{figure}

In this section we describe the main hardware parts that have been upgraded.
The individual hardware items of the upgrade program are shown in Fig.~\ref{fig:MAGIC-Scheme}.

\subsection{Camera of the MAGIC-I telescope}
\label{magic-camera}

\begin{figure}[hbt]
\centering
\includegraphics[width=1.0\linewidth]{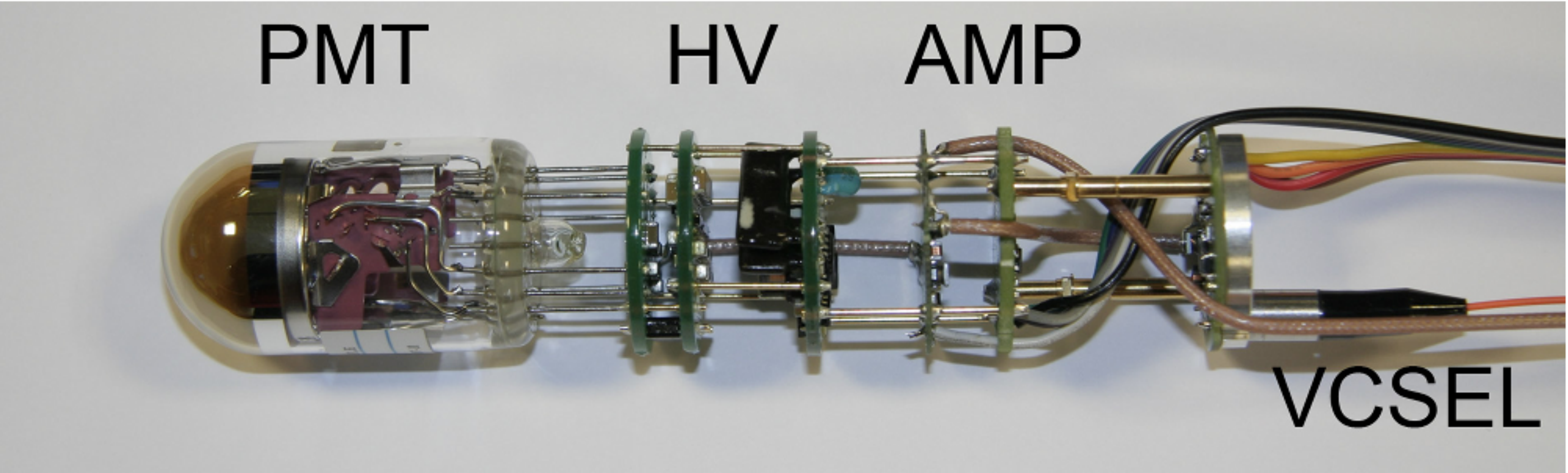}
\centering
\includegraphics[width=1.0\linewidth]{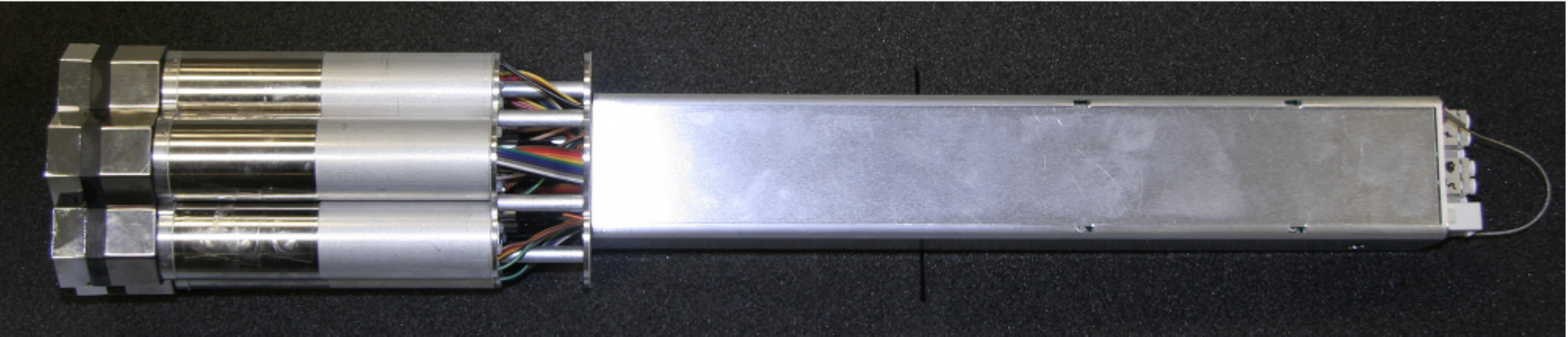}
\caption{\label{fig:pmt}
Assembled PMT module to form a pixel in the upper image and a full cluster of 7 pixels in the bottom image.
A PMT module includes the actual photomultiplier tube (Hamamatsu R10408),
its own HV generator, a preamplifier 
and a VCSEL, which transmits the analog signals to
the multi-mode optical fiber.  
{\bf As a safety measure, a diode is used to protect the amplifier input against too high current spikes due to strong illuminations.}
} 
\end{figure}

The new MAGIC-I camera has 1039 channels and follows closely the design and the performance of the MAGIC-II camera \citep{borla-tridon:2009a}.
The photosensors are photomultiplier tubes (PMTs) from Hamamatsu, type R10408, 25.4\,mm diameter, with a hemispherical photocathode and 6 dynodes, with an hexagonal shape Winston cone mounted on top. 
Each pixel module includes a compact power unit providing the bias voltages for the PMT and a stack of round circuit boards for the front-end analog signal processing, see the configuration in the upper photo of Fig.~\ref{fig:pmt}. 
The PMT bias voltages for the cathode and dynodes are generated by a low power, nine step Cockroft-Walton DC-DC converter, which can provide up to 1250\,V peak voltage. 
The electrical signals are amplified (AC coupled, $\sim$\,-25\,dB amplification) 
and then transmitted via independent optical fibers (no multiplexing) by means of vertical cavity surface emitting lasers (VCSELs).
The average pulse width signal is measured to be 2.5\,ns (FWHM) \citep{borla-tridon:2009a}.
{\bf The pixels are grouped in clusters of 7 to form a modular unit for an easier installation and access for maintenance}
(lower picture in Fig.~\ref{fig:pmt}).
A single cluster weighs around 1\,kg, has a length of 50\,cm and {\bf a} width of 9\,cm, with the distance of 3\,cm between the pixel centers.
We operate the PMTs at a rather low gain of typically $(3-4)\cdot10^{4}$ (see below) in order to also allow observations under moderate moonlight without damaging the dynodes.
An electrical signal (called pulse injection) can be injected at the PMT base of every pixel allowing for daytime tests of the whole electrical chain from the PMT base down to the readout and trigger without applying a high voltage to the PMTs. 
The pulse injection signals have similar shape as the Cherenkov light pulses (FWHM of 2.6\,ns) to have a realistic system response.
The amplitude of the pulses is stable over time and can be tuned from tens of photoelectrons up to saturation by means of two adjustable attenuators. 
The time jitter is of the order of 1\,ns.

The PMT gain is the main difference in the pixels from the upgraded MAGIC-I camera with respect to those from MAGIC-II.
The gain distribution
of the PMTs for the MAGIC-II camera goes from $1.0\times10^4$ to $6.0\times10^4$
with the mean at $3.0\times10^4$ (all measured at 850\,V). Such a spread is typical for PMT gains,
and the costs to make the gain distribution narrower would {\bf have} been disproportionally
high.  The differences in gain for different PMTs are compensated by adjusting
the high voltage (HV) settings of the PMTs independently with the so-called
flatfielding procedure (see Section~\ref{hvflatfielding}). This leads to a
significant spread of applied HVs. During the operation of MAGIC-II it proved
to be more practical to operate PMTs at a higher gain, typically $\sim4\times10^4$.
This increased the signal to electronic noise ratio and helped in the low
$\gamma$-ray energy analysis.  However, for such target gain, some PMTs had to be operated at the
highest possible voltage, and their number was increasing with time due to aging
effects. Therefore, when procuring PMTs for the MAGIC-I upgrade camera, it was
decided to order half of the PMTs with a higher gain: $1.5\times10^4$ to
$9.0\times10^4$ with the mean at $4.5\times10^4$. After purchasing, the PMTs
were selected in ``high-" and ``low-gain" ones according to their actual gain by
making a cut at $3.0\times10^4$.
The analog signals of the ``high-gain" 
PMTs are then attenuated in the PMT clusters by a factor of two (using a resistor), 
resulting in an overall narrower gain distribution of the PMTs in the MAGIC-I camera, see Section~\ref{hvflatfielding} for details.   
In the same time 69\,PMTs in the MAGIC-II camera showing the lowest gain were replaced
by PMTs with a higher gain, which allowed to {\bf minimize} the number of PMTs operated at the maximum HV.

\subsection{Optical cables}

Optical cables 
continuously transmit analog signals from the PMTs to the readout and trigger electronics located in the control house.  
The optical fibers are $\sim$162\,m long and are grouped in 19 bundles (per telescope) for a better handling, 72 fibers each, allowing for sufficient amount of spare fibers in case some break.
The bundles are protected by a UV resistant PVC cover to ensure mechanical rigidity, protect the fibers from breaking and from the strong sun UV radiation in La Palma.
It is important to prevent divergence of arrival times between individual channels due to different times of flight in the optical fibers.
Therefore, a special setup was developed to control that the propagation time is uniform in the fibers. 
The resulting spread in the propagation time is 138\,ps (RMS), and maximum difference of 650\,ps. 
The spread in propagation time is important at the trigger level when combining signals from individual pixels to form a telescope trigger and for the timing parameters of the shower image after the signal extraction. 
The former time spread is corrected online (see Section~\ref{trig_adjustment}) and the latter one is corrected offline using reference calibration signals.
No environmental factors have been noticed to affect the propagation time
of the signals in the fibers.  The exchange of the previous MAGIC-I fibers was
necessary because of the high density of the channels at the {\bf new} receiver boards
(see Section~\ref{sec:receiverboards}) that required smaller optical connectors.

\subsection{DRS4 based readout}
\label{drs4section}

The DRS4 based readout system is the major technical novelty of the upgrade.
The baseline concept of the readout system, now adopted in both telescopes, is the same as the one used in MAGIC-II in 2009 \citep{tescaro:2009a}. 
The readout electronics is divided in two main parts: the receiver boards and the
digitization electronics, both controlled by the same VME-based communication
network\footnote{CAEN-CONet daisy-chain network (using the CAEN A2818 PCI-card
and the CAEN V2718 optical linked VME bridges).}.

\begin{figure}[t]
\begin{center}
\includegraphics[width=0.85\linewidth]{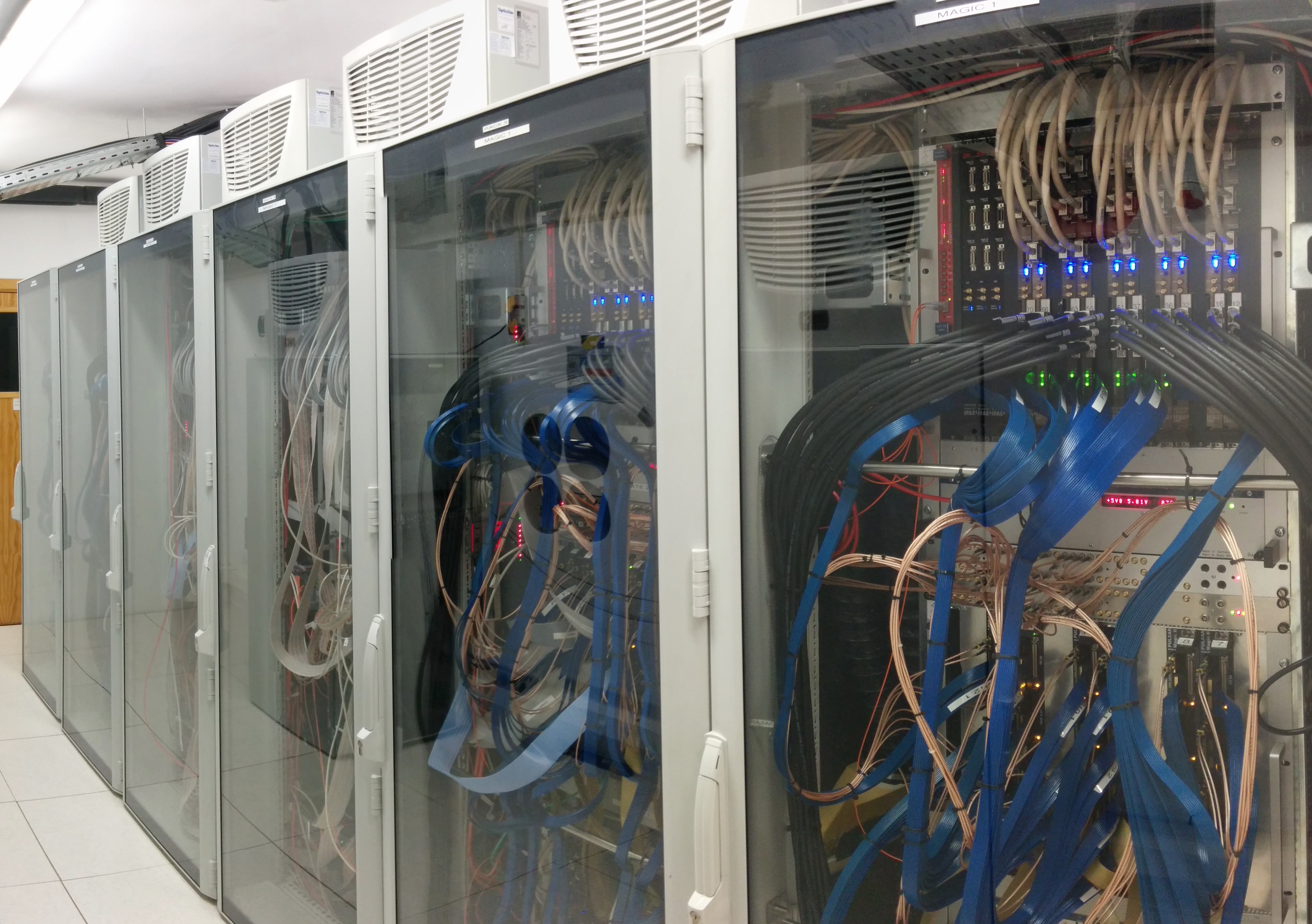}
\caption{
View of the electronics room of the MAGIC telescopes.
The six closed racks can be seen. They are placed on a technical raised floor (20\,cm height) allowing for better cable routing.}
\label{fig:ElectronicRoom}
\end{center}
\end{figure}

Cherenkov flashes last few ns only. To increase the S/N ratio and effectively exploit the arrival time information a fast sampling speed is needed (the time resolution goes roughly as 1/speed).
The new MAGIC readout {\bf is} sampling the signals with 2\,Gsamples/s. {\bf It} is cost effective, has a linear behavior over a large dynamic range (from less than 1 photoelectron (phe) to about 600\,phe), 
less than 1\% dead time, low noise, and negligible channel-to-channel cross-talk \citep{sitarek:2013a,bitossi:2014a}. 
This allowed us to maintain the performance of the previous readout based on MUX-FADCs while increasing the charge resolution, reducing cost and saving space. 
Reducing the space occupied by the readout electronics was very important.
In fact, the electronics room hosting the trigger and readout of the two telescopes was not large enough to host a readout of more than 2000 channels in a previous configuration.
Through the upgrade to a more compact DRS4 system (96 readout channels per 9U board), only 6 racks are needed for the trigger and readout system of the two telescopes (see Fig.~\ref{fig:ElectronicRoom}).

\subsubsection{Receiver boards}
\label{sec:receiverboards}

PMT signals are split in the Magic Optical Nano-Second Trigger and Event Receiver (MONSTER or receiver boards in short) into analog -- readout and sum-trigger, see below -- and digital branches. 
The optical fibers, carrying the optical PMTs signal to the control house, connect on the back side of the MONSTER boards by means of LX5-LX5 optical connectors. 
The MONSTER is a multilayer 9U board with the following tasks:
\begin{itemize}
 \item convert optical signals from the camera back to analog electrical ones;
 \item bring analog signals to the digitization electronics;
 \item generate the Level-0 (L0) individual pixel trigger signal using discriminators;
 \item further split the analog branch in order to feed a copy of the signals to the analog trigger
(sum-trigger, \cite{jezabel:sumtrigger:icrc:2013a});
\end{itemize} 

In the analog branch, the optical receivers have a bandwidth of 800\,MHz, a gain of 18.5 dB,
a negligible cross-talk of 0.1\% and a working range from 0.25 mV (corresponding to $\sim$ 0.15\,phe) to 1150 mV, with an RMS noise smaller than 0.2\,mV. A single board holds
24 channels with a maximum power consumption of 75\,W.

Three parameters of the L0 trigger can be adjusted from a PC via VME for each individual channel: \emph{(a)} the discriminator thresholds (DT), \emph{(b)} the delay, and \emph{(c)} the width of the output pulse of the discriminators. 
The thresholds and the widths/delays can be adjusted with a precision of 0.07\,mV ($\sim$ 0.04\,phe), and 10\,ps, respectively.
The individual pixel rate (IPR) can be monitored at a rate up to 1\,kHz but is currently monitored at 1\,Hz, which is sufficient for a reaction to stars in different fields of view (see Section~\ref{sec:IPRC}).

\subsubsection{Digitization electronics}
\begin{figure}[htb] \centerline{
\includegraphics[width=.9\linewidth]{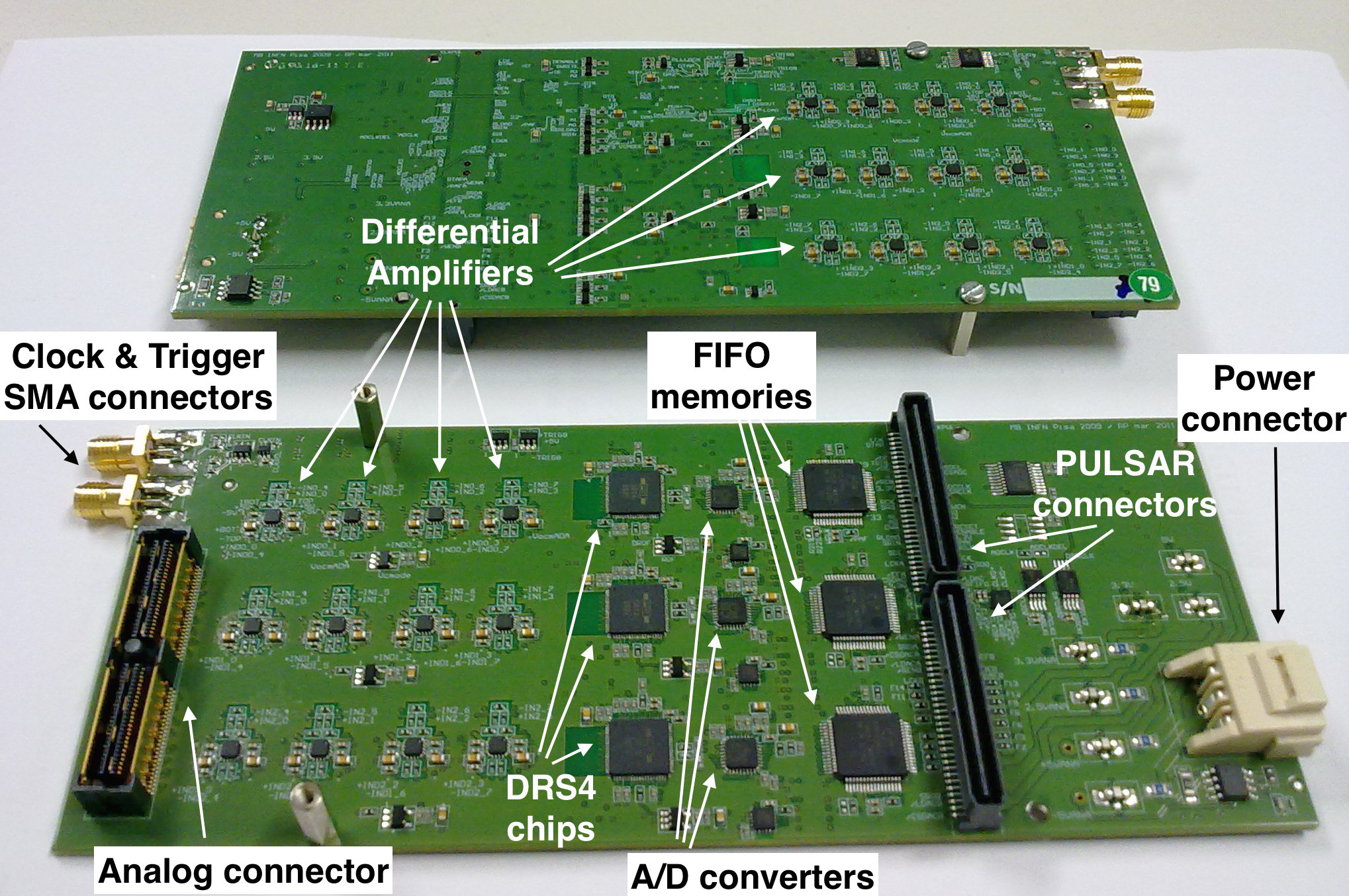}}
\caption{Picture of the DRS4 mezzanine developed at the INFN/Pisa electronics laboratories. 
From left to right one can recognize the two SMA connectors for the external synchronization signals (trigger and reference clock), the analog connector, the operational amplifiers to drive the input signal to the DRS4 chips, the three DRS4 chips (using 8 digitization channels each, 24 channels in total per mezzanine), the three built-in FIFO memories, the connector to interface the host motherboard and the external power supply connector.} 
\label{picdrs4mezz} 
\end{figure}

The sampling electronics is built with a motherboard-mezzanine logic, where the motherboard is the PULSAR board designed at the University of Chicago\footnote{http://hep.uchicago.edu/\~thliu/projects/Pulsar/.}, and the mezzanine is the new DRS4 mezzanine (Fig.~\ref{picdrs4mezz}) designed at the INFN/Pisa laboratory \citep{bitossi:2014a}.  
As mentioned above, the new DRS4 mezzanine uses now the DRS4 chip instead of the DRS2 chip adopted in 2009 for MAGIC-II. 
DRS4 stands for Domino Ring Sampler version 4, to distinguish it from its predecessor DRS2. 
We kept the same motherboards after a proper FPGA reprogramming. In fact, in the new version a single PULSAR board hosts 96 readout channels,
whereas in the DRS2 version it hosted 80 channels only.
Conceptually, it is an ultra-fast analog memory (a ring buffer built of 1024 switching capacitors) that is read out -- only in the event of a trigger -- at a lower speed by a conventional analog to digital converter.
In our case we use a 14-bit nominal resolution analog to digital converters (ADC), clocked at 32\,MHz. 
The raw pedestal level is set to $\sim$2500 ADC counts to allow the sampling of the negative part of the signals (like NSB fluctuations or pulses undershoots).
The DRS4 chips have a built-in Region of Interest (RoI) selection mode that reduces drastically the time overhead for the readout of the chip. 
The total dead time is dominated by the readout time of the DRS4 chips and is measured to be 27\,$\mu$s only (negligible in standard data acquisition conditions).  
The DRS4 chip has {\bf a} tunable sampling frequency (from 700\,Msamples/s to 5\,Gsamples/s) set to 2 Gsamples/s and {\bf a} linear response in an input range of 1\,V.
The mezzanine noise is $\sim$7.5\,ADC counts, corresponding to $\sim$450\,$\mu$V at the board input, and is dominated by the noise from the DRS4 chip which varies up to 50\% from chip to chip \citep{bitossi:2014a}. 
The measured bandwidth is $\sim$\,650\,MHz.  
Overall, the digitization electronics contribute to $\sim$50\% of the total noise (see Section~\ref{sec:sourcesofnoise}).

The DRS4 mezzanines (hosted by the PULSAR motherboards in groups of four) are connected to the receiver boards by means of  24 differential lines analog cables and synchronized by two SMA cables (one for the trigger signal and one for the common reference clock signal).
A total of 48 DRS4 mezzanines are installed in each readout, for a total of 1152 readable channels (enough to cover the 1039 camera pixels and keep $\sim$10\% spare channels).  

%

The final data acquisition (DAQ) is performed in a single computer per telescope steered by a multithread C++ program \citep{tescaro:2013a}. 
The readout electronics communicates with the DAQ via the SLink
optical data transfer system, with the HOLA cards attached on readout side and the FILAR PCI cards on the computer side\footnote{
SLink is a high speed (160\,MB/s) data transfer link with HOLA as sender and FILAR as receiver cards, 
all developed for the LHC experiments at CERN. See: https://hsi.web.cern.ch/hsi/s-link/ for more information.}.

\subsubsection{Readout data pre-processing}
\label{pre-processing}

The calibration of the chip response is mandatory to obtain optimal results in terms of noise and time resolution \citep[see][]{sitarek:2013a}.  
Three important corrections are applied to the data:
\begin{itemize}
\item The mean cell offset calibration;
\item The readout time lapse correction;
\item The signal arrival time calibration;
\end{itemize} 
Currently the first two are applied online by the DAQ program whereas the third is applied offline (although all the corrections can be applied offline if required).

\begin{figure}[htb]
\begin{center}
\includegraphics[width=1.0\linewidth]{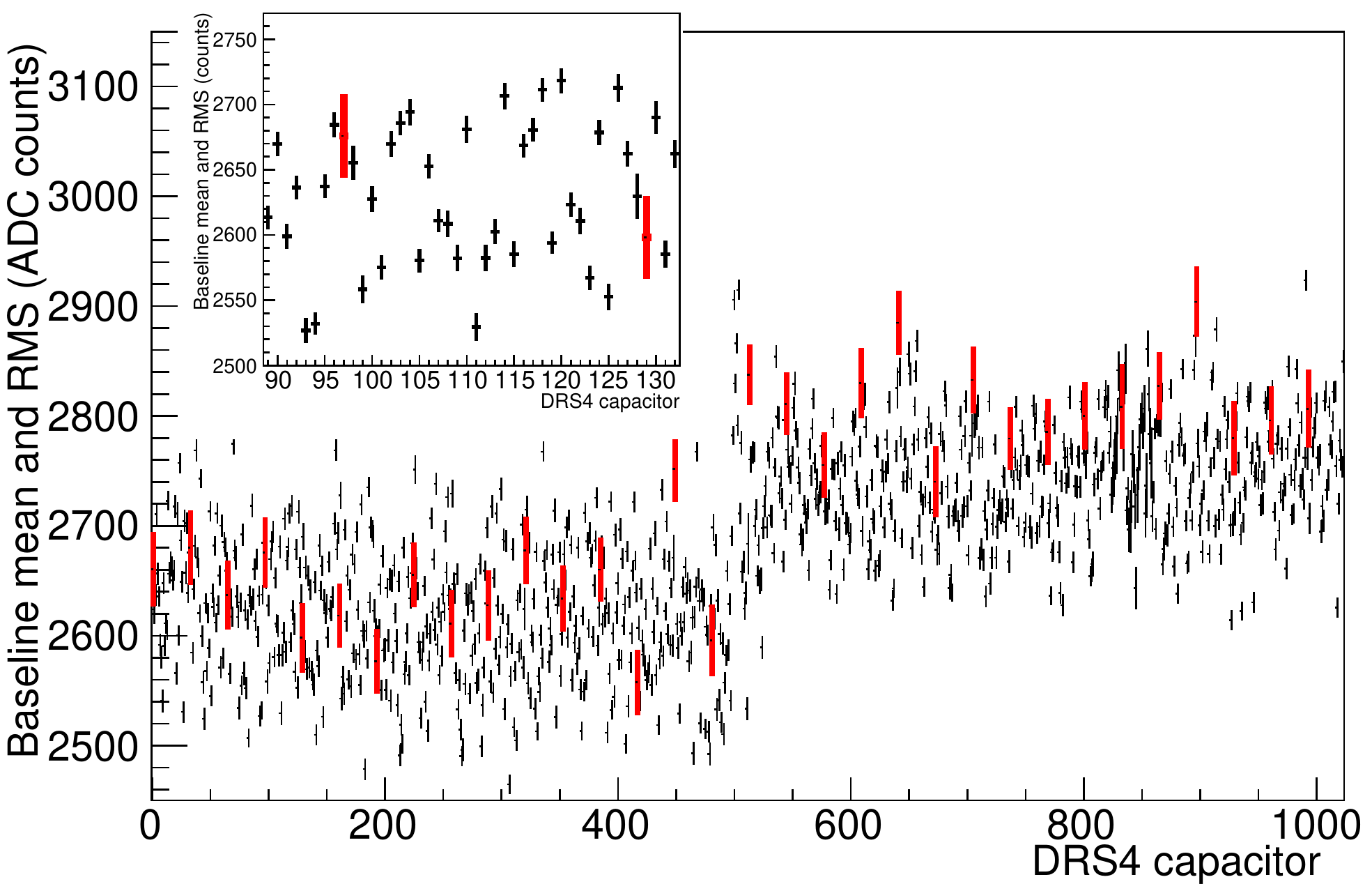}
\caption{
Cell offset of 1024 individual capacitors of one channel of the DRS4 chip. 
Vertical error bars show the standard deviations of the offset values for the capacitors. 
Every 32$^{nd}$ capacitor is marked with a thick red line. 
The inside panel zooms into some of the capacitors to better appreciate the differences from capacitor to capacitor \citep{sitarek:2013a}.
}
\label{fig:baseline}
\end{center}
\end{figure}

The mean cell offset is defined as the raw mean ADC count value for a certain capacitor during a pedestal run.
Fig.~\ref{fig:baseline} shows the mean cell offset (and its RMS) as a function of the absolute position of the capacitor (cell units) in the DRS4 ring buffer for a typical channel.
Notice that the single capacitor baseline varies up to $\sim$\,15\% from cell to cell, well beyond the noise fluctuations. 
To equalize the response and obtain a flat baseline the mean cell offset of each cell is computed using a dedicated DRS4 pedestal calibration run (taken once at the beginning of the night), and subtracted to the readout values.
This is what we call the mean cell offset calibration of the chip and has to be done with a special algorithm that takes into account not only the absolute capacitor position in the buffer but also the trigger position in the ring  \citep[see][]{sitarek:2013a}.

\begin{figure}[htb]
\begin{center}
\includegraphics[width=0.94\linewidth]{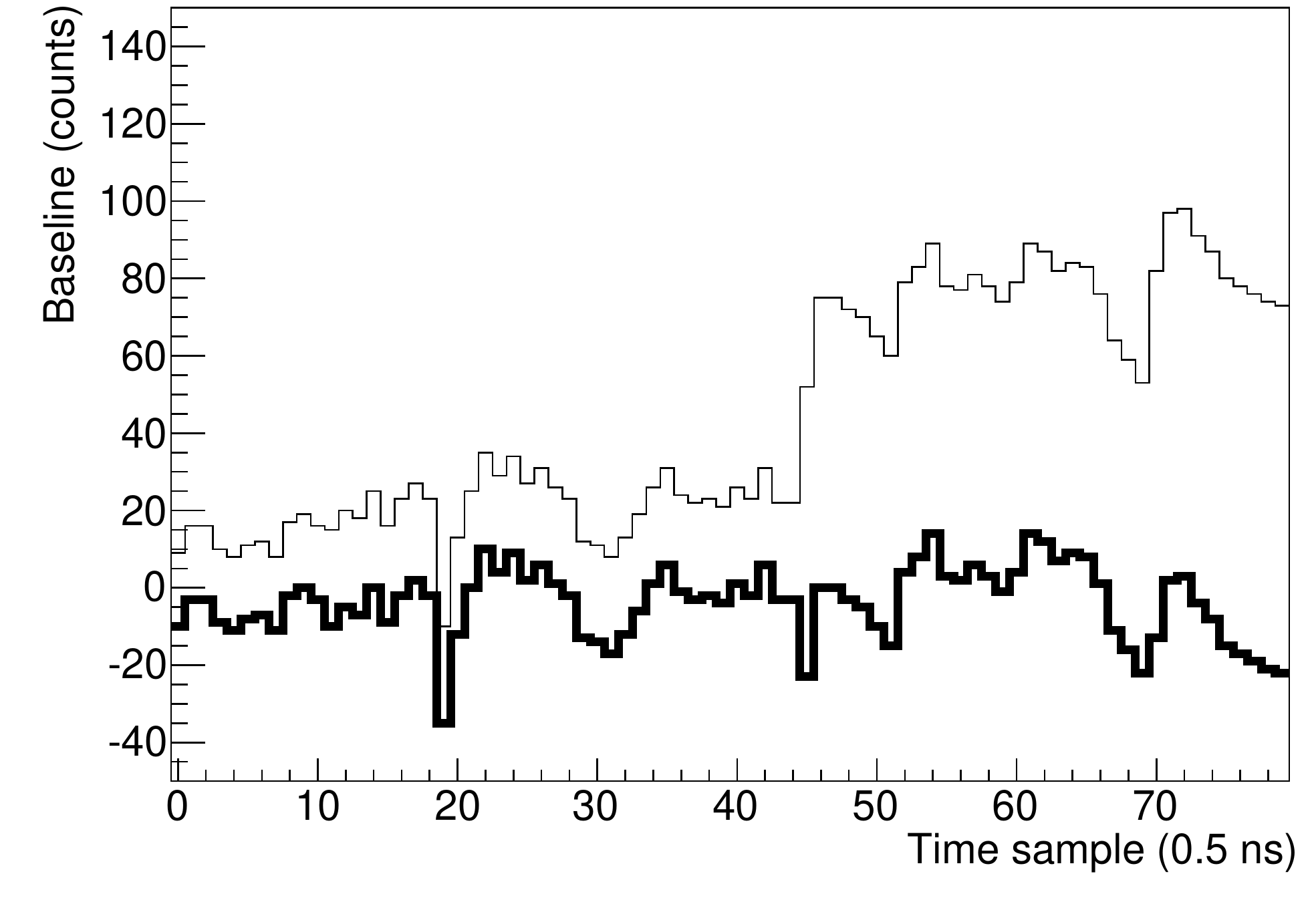}
\caption{
Example of digitized pedestal data with the DRS4 chip. The capacitor offset depends on the time lapse with the last readout of the capacitor, which results in steps on the baseline in case of 
non-fixed frequency triggers (like the cosmic-ray triggers, which follow Poisson distribution).
The thin line shows the original data and the thick line shows the effect of the time lapse correction, which recovers to a flat baseline. 
In this example the first half of the range needs only a small correction whereas the second half, that presents a clear step, requires a larger correction.
}
\label{fig:baselinejump}
\end{center}
\end{figure}

The mean cell offset calibration has to be further corrected since the mean offset suffers a dependency with respect to the time passed since the last reading of the cell: the offset decreases following a simple power law as a function of the time lapse.
Since this behavior is very similar for all the DRS4 chips, a universal analytical expression can be used to further correct the single capacitor's offsets. 
If not corrected, this effect would produce steps in the baselines (see Fig. \ref{fig:baselinejump}), since for a given readout cycle of the chip only a small part of the buffer is actually readout.

\begin{figure}[htb]
\begin{center}
\includegraphics[width=1.\linewidth]{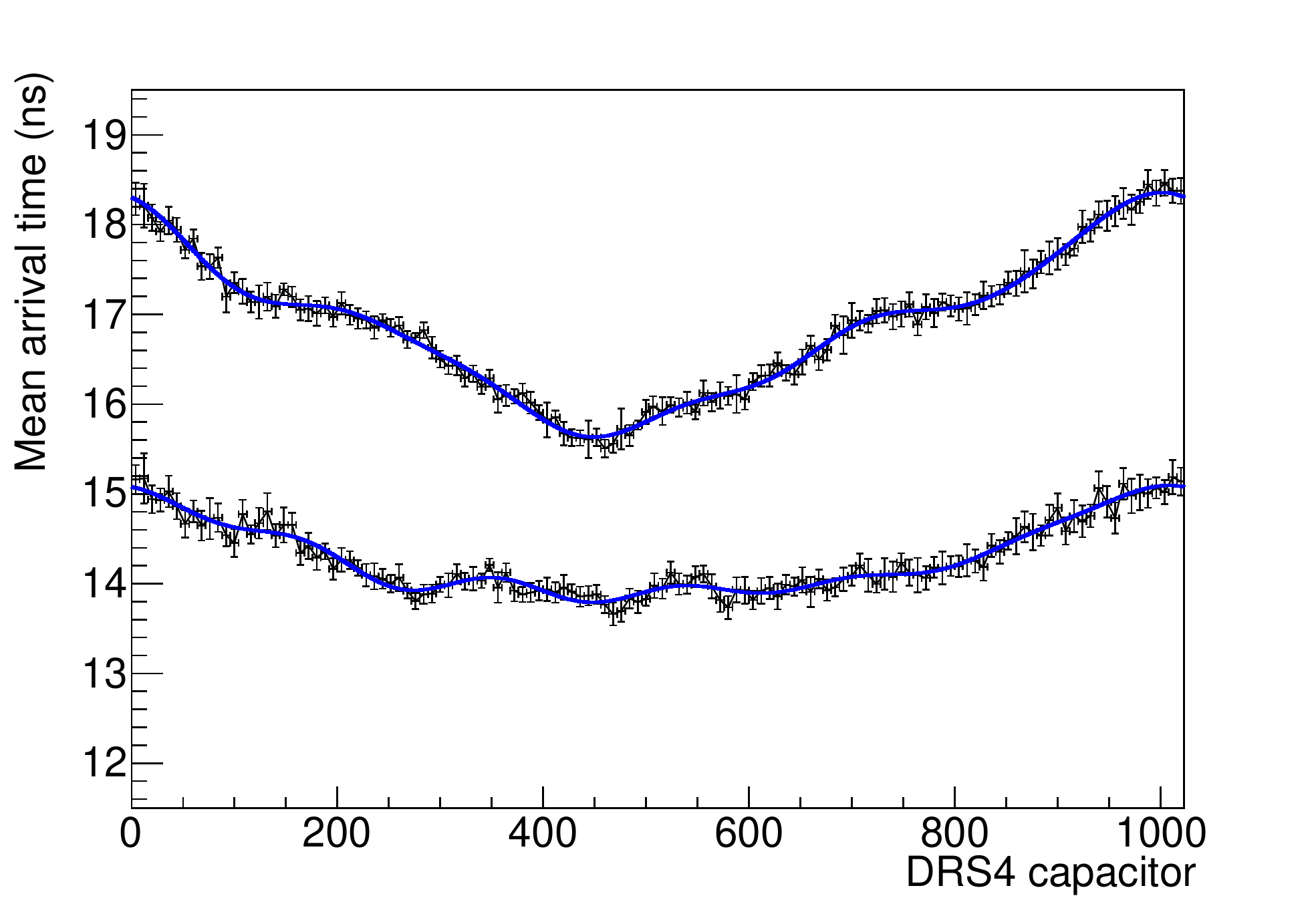}
\caption{
Mean pulse signal arrival time as a function of the position in the DRS4 chip run buffer for {\bf two} typical channels together with their Fourier series expansion (solid lines).
}
\label{fig:timecalib}
\end{center}
\end{figure}

Finally, similarly to the DRS2, DRS4 channels exhibit a moderately variable time spread (1--4\,ns) on the delay of the recorded signal pulses, depending on the absolute position in the ring buffer (see Fig. \ref{fig:timecalib}).
This effect is chip-dependent and has to be calibrated independently for each DRS4.
The characteristic delay figures are built by means of calibration runs (synchronous pulses of fixed amplitude) and parameterized using Fourier series expansions.
This basic arrival time calibration recovers the true arrival time at the DRS4 input, resulting in a characteristic time spread of $\sim$\,0.2\,ns \citep{sitarek:2013a}.
The normal calibration runs taken during data taking (several per night) are used for this purpose.

\subsection{Individual telescope trigger and stereo trigger}
\label{magic-trigger}

In the MAGIC-II camera and the upgraded MAGIC-I camera the trigger region covers the 547 inner pixels. 
The MAGIC trigger has three levels.
The first trigger level (L0) is a simple amplitude discriminator operating on each pixel individually. 
For each telescope, the 547 digital L0 signals generated by the receiver boards (see
Section~\ref{sec:receiverboards}) are sent to the second trigger level, the telescope trigger (L1).
%

\begin{figure}[htb]
\begin{center}
\includegraphics[width=0.9\linewidth]{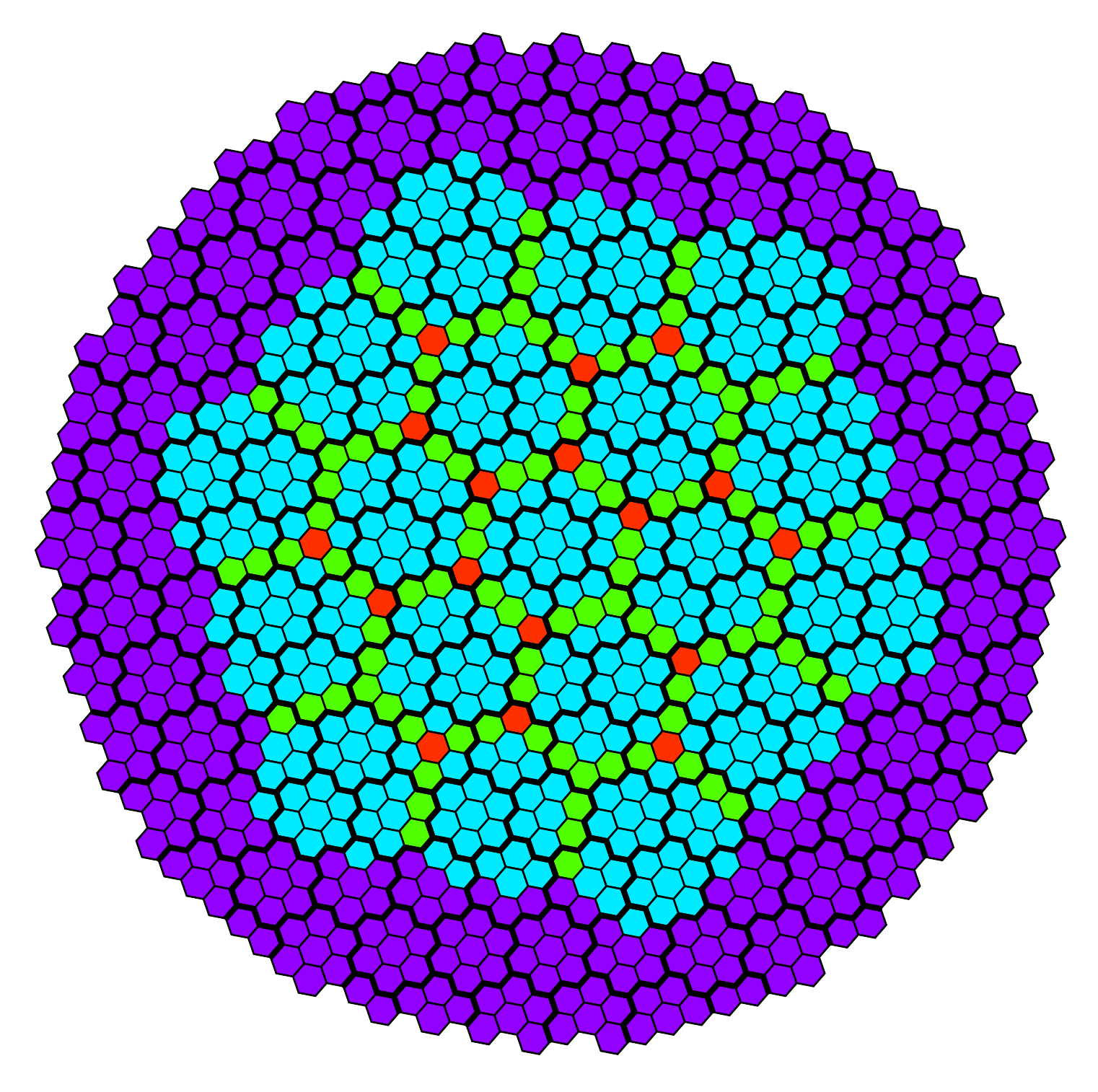}
\caption{
Geometry of the MAGIC cameras with 1039 channels, each. 
The cyan hexagons (36 pixels each) show the 19 L1 trigger macrocells. 
Pixels that are covered by more than one macrocell are shown in green (single overlap) or red (double overlap).
The thick black lines marks the layouts of the cluster of pixels.
}
\label{fig:macrocells}
\end{center}
\end{figure}

The L1 trigger is a digital filter arranged in 19 macrocells of 36 channels each, with a partial overlap of channels between the macrocells (see Fig.~\ref{fig:macrocells}). 
Several logic patterns are implemented: 2 next-neighbor logic (2NN), 3NN, 4NN, and 5NN.
All patterns are close compact. Only one pattern logic can be selected at a time, which is done at the beginning of every observation. 
In case any of the 19 macrocells reports a coincidence trigger of the programmed logic, a L1 trigger signal (also called individual telescope trigger) is issued.  
The upgraded trigger for MAGIC-I has the same number of macrocells as the previous one (and the one of MAGIC-II) but the overlap between them was reduced from three pixel rows to one, since the number of trigger circuits did not increase.
The smaller overlap created $\sim$1$\%$ trigger inefficiency over the field of view for 3NN and 4NN logics but increased the trigger area by a factor of $\sim$1.7 
with respect to that of the old camera.

The {\bf two} L1 trigger signals ({\bf one per camera} generated using the 3NN logic) are sent to the third trigger level, the stereo trigger (L3).
The L1 signals are artificially stretched to 100\,ns width and delayed according to the zenith and azimuth orientation of the MAGIC telescopes to take into account the differences in the arrival times of the Cherenkov light from air showers 
at the corresponding focal planes.  
A logical {\bf 'or' operation is made} 
between the two signals, and the resulting signal (L3 output) is sent back to the individual telescope readout.
The width of 100\,ns for the two signals is chosen to ensure a safe margin for a 100\% L3 efficiency even in case of some misalignment in the timing between the two telescopes. 
The L3 coincidence has an intrinsic jitter of about $\pm10$\,ns due to the angular difference between the shower axis of the triggered events and the pointing direction of the telescopes.
The maximal delay between the two L1 signals capable to produce an L3 trigger is $\sim$\,200\,ns.
We describe performance parameters of the trigger in the commissioning Section~\ref{sec:commissioning}.

\subsection{Calibration system}
\label{MAGICCalibSystem}

The calibration of the MAGIC telescopes is performed through the uniform illumination of the PMT camera with well-characterized light pulses of different intensity produced by a
system, which we name \emph{calibration box}, installed at the (approximate) center of the mirror dish, i.e.\ about 17\,m away from the camera plane. 
The \mbox{MAGIC-I} calibration box was installed in 2004, and was based on fast-emitting (3-4 ns FWHM) LEDs \citep{schweizer:2002a}.
The light intensity was adjusted by changing the number of LEDs that fired, and the uniformity was achieved by a diffusor at the exit window. 
On the other hand, the MAGIC-II calibration box (installed in 2009) is based on a system with a
passively Q-switched Nd:YAG laser (third harmonics, wavelength of 355\,nm) that produces pulses of 0.4\,ns FWHM.
The light intensity is adjusted through the selection of a calibrated optical filter
and the uniformity is achieved by means of an Ulbricht sphere that diffuses the light right before the exit window.
After the Ulbricht sphere the laser pulse has a FWHM of $\sim$1\,ns, which is similar to the time spread of the photons in the Cherenkov shower 
\citep{aliu:2009a:timinganalysis}.

The laser-based system was proven superior to the LED-based system because it provides \emph{(a)} 
a larger dynamic range, and \emph{(b)} shorter light pulses ($<2$\,ns FWHM), which are more similar to the ones produced by the Cherenkov flashes from extended air showers.
For the upgraded MAGIC system (both MAGIC-I and MAGIC-II) we decided to use a calibration box similar to that originally installed in MAGIC-II but with some performance upgrades:
\emph{(a)} a humidity sensor inside the box, \emph{(b)} the laser status can now be queried, 
\emph{(c)} a heating system attached to the entrance window to avoid water condensation, \emph{(d)} a fast photodiode for monitoring the laser light output, \emph{(e)} an improved dynamic range, together with a more detailed characterization of the light intensities, and \emph{(f)} an improved uniformity in the illumination of the telescope camera with variations of less than 2\%\footnote{
The homogeneity was evaluated in the lab, with a PMT matrix located at a distance of 4\,m from the calibration box, and a computer-controllable
turning table that rotates the calibration box (with the rotation axis going through the Ulbricht sphere) from -5 to +5 degrees in steps of 0.05 degrees (half the size of one pixel). 
}

Before the observation of a new source, a calibration run consisting of 2000 events at a fixed light intensity is taken. 
The extracted charge per pixel and its variance are used to determine the conversion factor between the ADC counts of the readout and the number of phe via the F-factor method, which relies on the knowledge of the added noise of the PMT \citep{mirzoyan:1997a,schweizer:2002a}. 

The temporal stability in the illumination of the camera is given by the temporal stability of emitted laser light, which is better than 1\% for (short) timescales $\sim$10\,min, and better than 5\% for timescales of days (as tested in the laboratory). 
The emitted laser light should be stable over months and years timescales, until the aging of the crystal starts taking place, which nominally occurs only after 5000\,h of operation (accordingly to specifications from the manufacture).
In any case, the calibration of the PMT signals in MAGIC is possible even if the laser light drifts over time. 
This is due to the fact that the calibration system is used to obtain the conversion factors between input (number of phe produced in the photocathode and collected by the first dynode of the PMTs) and output (measured number of ADC counts from the digitized signal), 
{\bf and the derived conversion factor should be correct provided that the laser light intensity does not change significantly on timescales of 10\,min, which is the timescale that is used to determine the input signal
 in phe through the F-factor method.}

The calibration light pulses are also used to cross-calibrate the analog arrival times in the DRS4 channels, which are different channel by channel (due to differences in propagation time
between the focal plane and the DRS4 chip) and depend on the position of trigger signal in the
DRS4 ring buffer \citep[see][]{sitarek:2013a}.
In addition, during data taking the calibration laser is constantly firing at 25\,Hz (so-called interleaved calibration events) allowing to monitor the gain in the readout chain of the individual channels.
The calibration system is also used for the fine tuning of the trigger signal delays described in Section~\ref{L0adjustement}.

\subsection{Computing} \label{computing}

The computing infrastructure of the MAGIC telescopes was also upgraded as a part of the general hardware upgrade.
Most of the computing equipment was moved from the electronics room to an adjacent, newly prepared dedicated computer room. 
Four racks containing computers, storage elements and network equipment were installed in the new location and connected mainly via Gigabit Ethernet but also 
via Fibre Channel in some special cases (DAQ computers to storage disks).
All equipment was connected to power switches that can be controlled remotely. 
New computers were also added to the cluster of analysis machines to process data on-site and the volume of the storage elements was doubled by adding new disks.
The computing system in La Palma is mainly a stand-alone cluster, connected through a gateway server to the external network.
Moreover, a dedicated machine is connected to the European Grid Infrastructure\footnote{http://www.egi.eu/} (``Grid" in the following) and is appointed to the data transfer to the MAGIC data center (see below). 
More details on the storage area configuration can be found in \citet{carmona:2009a}.

\begin{figure*}[phtb]
\begin{center}
\includegraphics[width=0.9\linewidth]{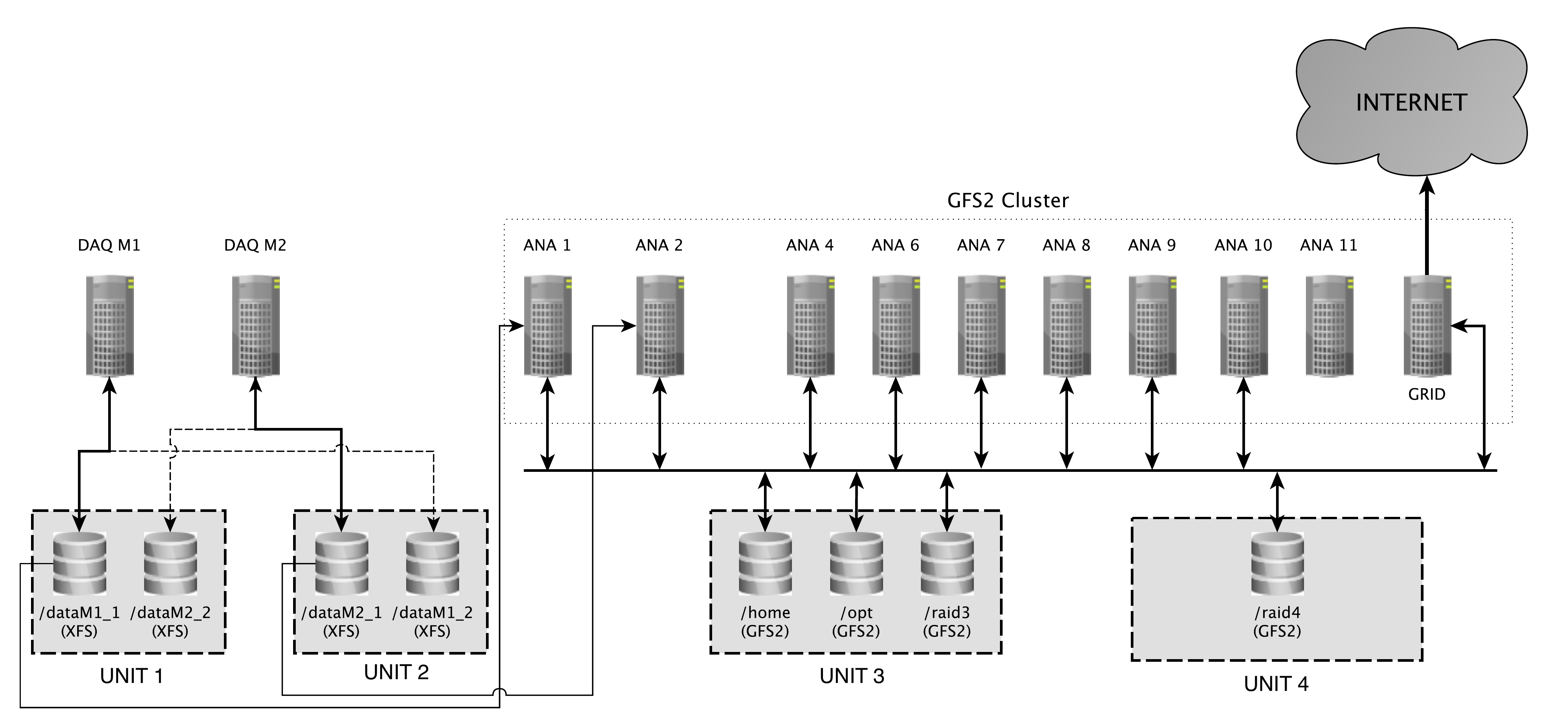}
\caption{
Schematic view on the computing network of the DAQ 
and the data analysis machines in MAGIC.}
\label{fig:MAGIC-Computing}
\end{center}
\end{figure*}

A major upgrade of the operating system became necessary since it was not possible to keep the old operating system for newer computers.
The computers are split into a cluster of the on-site analysis machines, subsystem machines (needed for operation of the telescopes) and the storage area network (SAN).
The analysis computers that can access the SAN were updated to a new operating system version (Scientific Linux CERN 6.3) and new computers were added. 
Two storage elements (Unit-1 and Unit-2), where the raw data is written by the DAQ machines, were separated from the rest of the GFS\footnote{Global File System for a shared disk file systems for Linux computer clusters.} cluster and formatted as XFS\footnote{XFS is a high-performance 64-bit journaling file system.}, see Fig.~\ref{fig:MAGIC-Computing}. 
The raw data is copied to Unit-3 (MAGIC-I data) and Unit-4 (MAGIC-II data) during the data taking. 
The raw compressed files are already available for full analysis on the Unit-3 and Unit-4 partitions few minutes after the end of the observations, which allows the on-site analysis 
{\bf machines} to start processing them timely.
The capacity in each of the two storage units, that are connected to the DAQ machines directly, is 7.3\,TBytes for the main partition and 3.7\,TBytes for the backup partition.
The volume in the main partition of this elements is large enough to contain more than 5 full nights of uncompressed data under normal operation conditions. 
The total capacity of the storage units is $\sim$100\,TB.

High level analysis results are produced on-site by all analysis nodes using
the standard analysis software MARS \citep{zanin:2013:magic:mars}, and are usually available a few hours
after the end of the data taking. 
In this way, the on-site analysis does not interfere with the data taking and raw file writing procedure.
These processed data files and the original raw data files are copied to the MAGIC datacenter at PIC (Port d'Informaci\'o Cient\'ifica, www.pic.es), in Barcelona, Spain \citep{reichardt:2009a}, where they are permanently stored and made available to the whole collaboration.
The copying of the data to PIC is done from the node of the GFS2\footnote{
GFS2 is a further development derived from GFS and was included along with 
its distributed lock manager (shared with GFS) in Linux 2.6.19.} cluster that is connected to Grid, using efficient Grid software tools for data transfer of large data volumes. 
The transferred data are accessible and analyzable by the users {\bf through} the Grid framework.
Fig.~\ref{fig:MAGIC-Computing} shows the schematic view of the configuration for DAQ machines and GFS2 cluster after the changes introduced in mid 2012. 
The configuration of the system is however very flexible and it allows to add easily new computers or new storage elements. 

\subsection{The timing system}

The timing system is used to {\bf deliver absolute time stamps for the acquired events at the moment when the trigger signal reaches the readout electronics.
The readout system itself is not synchronized to any external clock.}   
Before the upgrade the timing system consisted of several separate units: a Rubidium clock (Rubclock), a GPS module and several NIM modules.
The timing information were converted to low voltage differential signal (LVDS) format and fed to the readout at the time of the trigger in order to timestamp events, so the 44 bit LVDS signal required 88 physical cables divided into four connectors. 
The precision of the system was 200\,ns.
Since the system became difficult to maintain (several modules were in use since the HEGRA experiment in the 90's, \citet{daum:1997a}) it was decided to build a new timing system.

The Rubclock and GPS modules were substituted by a single commercial timing 
system\footnote{http://www.symmetricom.com/products/gps-solutions/gps-time-frequency-receivers/XLi/~~(Symmetricom XLi).}.
The system is coupled to a custom-built timing rack module, which contains all the electronics needed to export valid timing information in LVDS format for both MAGIC-I and MAGIC-II. 
The precision of the new system is the same as for the old one: 200\,ns.
However, in the old system there was a drift between the UTC 1 pulse per second (1PPS) and the 1PPS signal from the Rubidium clock, which could exceed 1.5$\mu$s and had to be reset manually, typically once a month.
The drift is reduced to 65\,ns/month in the new system and does not need a manual reset, which makes the system more accurate. 

\subsection{The central control program}

A central control program allows the telescope operators to perform and monitor observations 
\citep{giavitto:2013:thesis}. 
It allows to configure and control all subsystems of the telescopes. 
It provides a unified graphical user interface, which allows to easily execute many complex operations. 
Tasks that require the synchronization of many subsystems are coded as modular routines, which can be called individually by the user. 
All subsystem configuration parameters are kept in plain text configuration files. 
This architecture enables great flexibility and rapid development cycles. 
During the upgrade, the existing routines and configuration files were adjusted to the newly introduced subsystems. 
In some cases, the changes introduced by the upgrade permitted further automation of
some tasks, so the corresponding routines had to be coded anew. 
A real time monitor of the data readout was also written, allowing experts to inspect every channel down to the sample level.  
Many new features were included during the upgrade, these include: automatic startup and
shutdown procedures, an online monitor and long-term database of critical parameters of the telescopes (e.g., temperatures, rates etc.), and an automatic Gamma-Ray Burst pointing procedure not requiring human intervention\footnote{
Gamma-Ray Bursts are transient events of very short duration (the prompt phase lasts only a few seconds) and in order to increase chances to catch them with Cherenkov telescopes reaction time must be minimized. 
Therefore, automatic procedures not requiring human intervention are essential.}.
Furthermore, an automatic procedure has been established to routinely take images of stars for monitoring the telescope tracking accuracy, pointing precision and optical point spread function, PSF.

\section{Low level performance}

Here we shortly describe the basic performance parameters of the MAGIC telescope system after the upgrade.

\subsection{Sources of noise}
\label{sec:sourcesofnoise}

The two main sources of noise in the extracted signals are electronic noise and fluctuations of the night sky background (NSB). 
The goal of the upgrade was to keep the electronic noise at a similar level as the noise coming from the extragalactic (dark time, no bright stars) NSB.
The individual contributions of the noise were extracted by dedicated runs taken with certain
contributions on and off separately. First only readout electronics was switched on allowing to measure the contribution from the DRS4 and the receivers. 
Then the bias current of the camera VCSELs (see also in \citet{borla-tridon:2009a}) was turned on, and finally the HV was applied to the PMTs and camera opened during night pointing to a dark patch of the sky.
The assumption in determining the individual components of the electronic noise is that they are mainly independent of each other.
The obtained numbers are summarized in Table~\ref{tab:noise}.  
One can see that the electronics noise (RMS) from the readout is at the level of 0.7\,phe, the
contribution from the camera (mainly VCSEL for the optical signal transmission) of 0.3\,phe, which is to be compared with the level of the NSB of 0.6--0.7\,phe.
Note that the level of the electronics noise in phe depends on the target HV used in the flatfielding procedure (Section~\ref{hvflatfielding}).
The applied HVs to the PMTs do not contribute to the noise in any measurable way.
The measured NSB level is higher in MAGIC-II because of newer mirrors that have a higher absolute reflectivity than the MAGIC-I mirrors \citep{doro:2008:mirrors}.
The relative precision of the measurements is at the level of a few per cent. 
The absolute scale of the measurement is about 10\%, mainly due to the uncertainties
converting ADC counts into phe.



\begin{table}[t]
\centering
\begin{tabular}{c|cc}
Source               & MAGIC-I            & MAGIC-II \\ \hline
DRS4+receivers       & 0.76 phe & 0.69 phe\\ 
VCSEL                & 0.30  phe  & 0.30 phe\\ 
NSB (extragalactic) & 0.60 phe &  0.72 phe\\ \hline
Total & 1.0 phe  & 1.0 phe
\end{tabular}
\caption{
Contribution to noise from different hardware components as well as from the NSB for MAGIC-I pixels 
and MAGIC-II (in terms of pedestal RMS).}
\label{tab:noise}
\end{table}

\subsection{Linearity in the signal chain}

\begin{figure}[h!]
\begin{center}
\includegraphics[width=1.\linewidth]{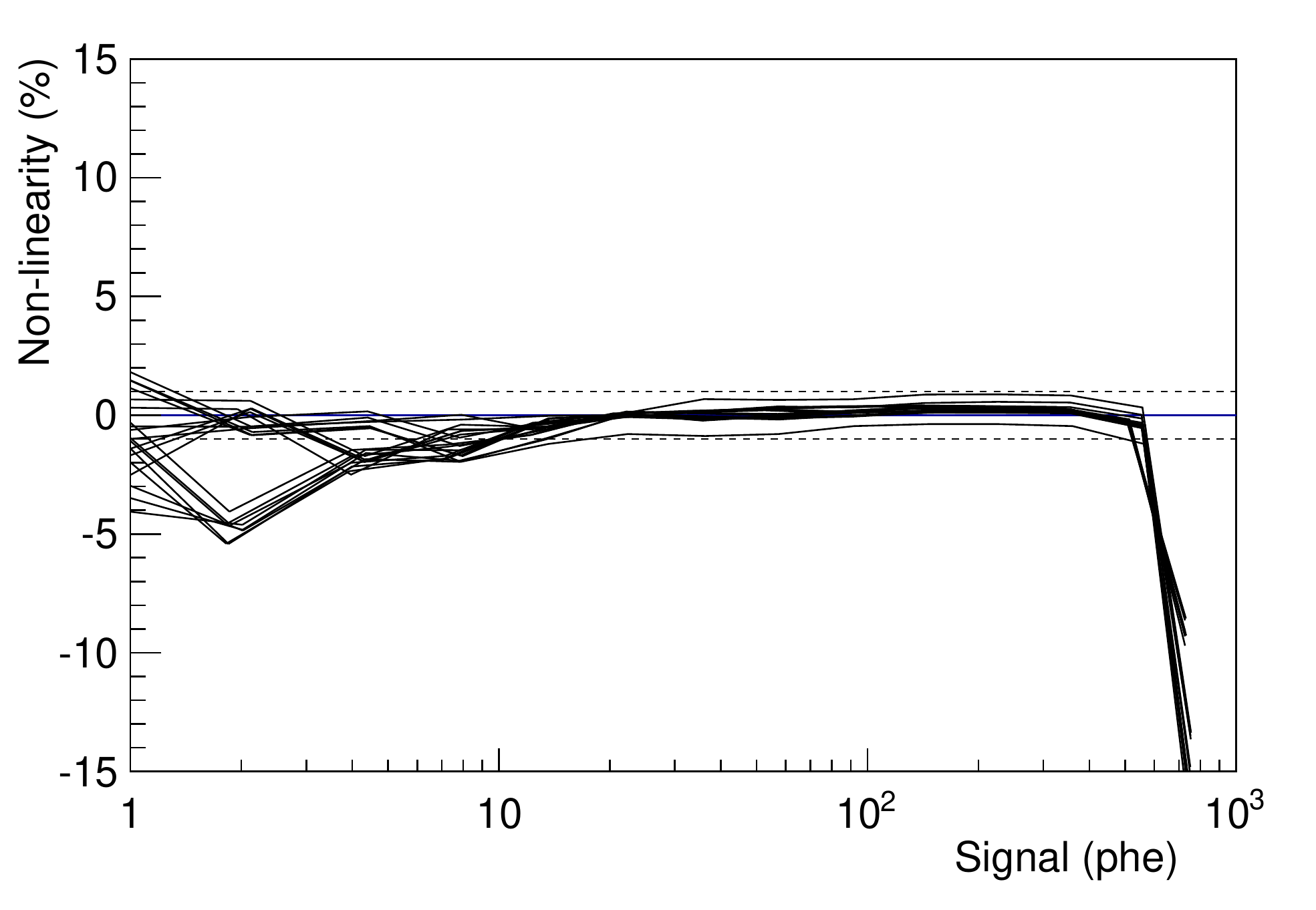}
\caption{
Deviation from linearity for 20 typical channels of the DRS4 readout \citep{sitarek:2013a}. 
The dashed lines mark 1\% deviation. 
A single photoelectron has an amplitude of $\sim$30 readout counts.
}
\label{fig:chainlinearity}
\end{center}
\end{figure}

For the linearity of the readout chain we refer to a more detailed study in \citet{sitarek:2013a}.  The linearity of the full electronics chain (PMT to the DRS4 readout) is better than 10\% deviation in the range from 1--2 phe (though it is very difficult to measure 1\,phe signals since the noise level is of the same order of magnitude) to few hundred phe (see Fig. \ref{fig:chainlinearity}). 
Some non-linearity of the order of 10-20\% is observed for pulses with charge between 200\,phe and 1000\,phe, and signals saturate the readout (at the stage of the receiver board) at $>$1000\,phe. 
The non-linearity effect at high charges is mainly due to the behavior of the VCSELs.
Simulations showed that a non-linearity of that magnitude does not affect image parameters of events with a charge lower than 10,000\,phe and has a 1--3\% effect for events with a higher charge, so that no linearity correction is required.

\section{Commissioning of the system}
\label{sec:commissioning}

The key point of the efficient commissioning was to have a dedicated and
well experienced team of 5 to 10 physicists at the site of the experiment for
a duration of several months after the installation of the hardware.  In the
following the main milestones of the commissioning are described.

\begin{figure*}[ht!]
\begin{center}
\includegraphics[width=0.49\linewidth]{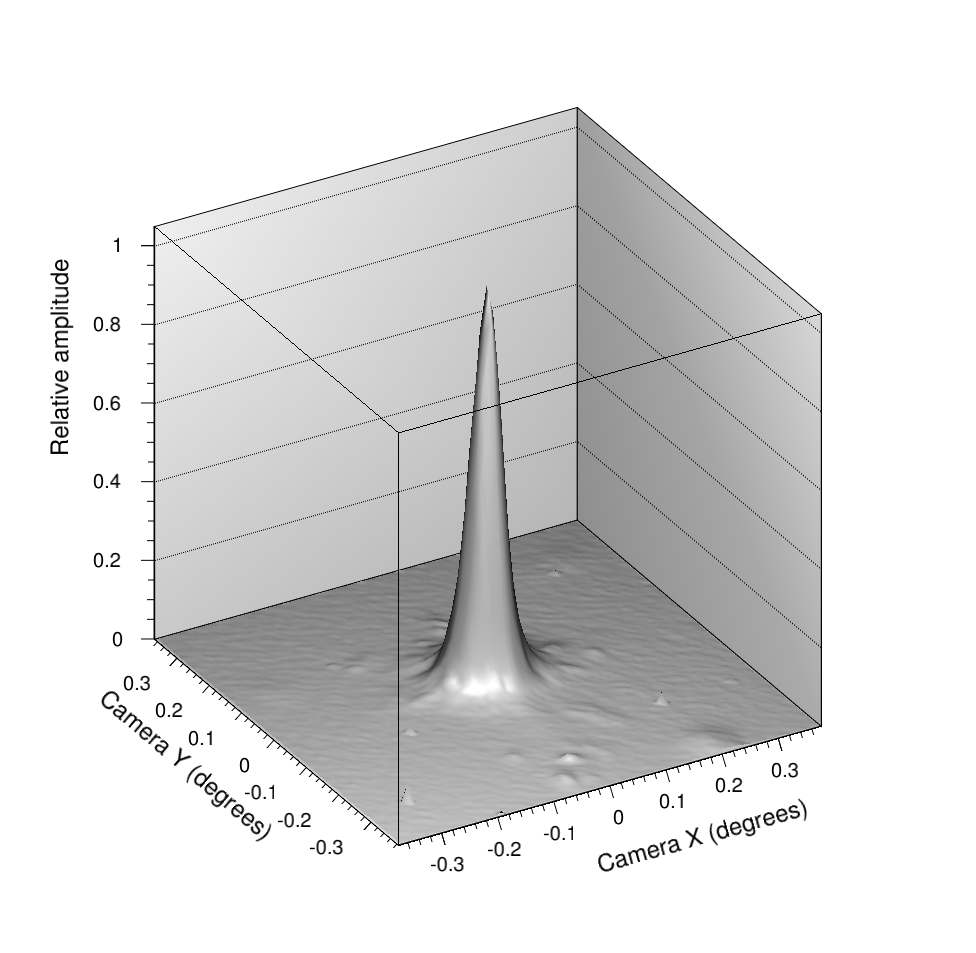}
\includegraphics[width=0.49\linewidth]{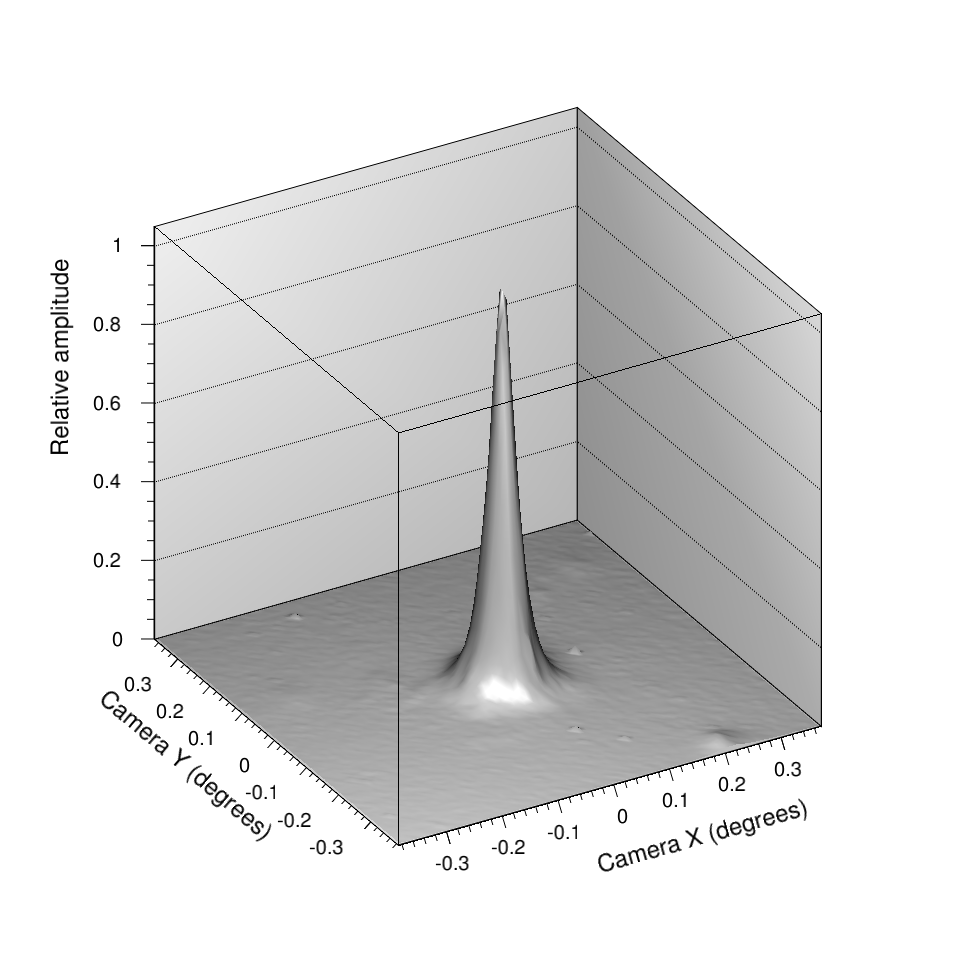}
\caption{
Optical point spread function for the two MAGIC telescopes (MAGIC-I left, MAGIC-II right).
The image of the star called Menkalinan taken with the SBIG$^{\textregistered}$ camera 
at a zenith distance of 16 degrees.}
\label{fig:PSF}
\end{center}
\end{figure*}


\subsection{Optical point spread function}

The optical point spread function (PSF) was improved during the upgrade.  
A dedicated active mirror control (AMC) hardware and software \citep{biland:2008a} takes care of mirror adjustment depending of the zenith angle of observation due to small deformations of the telescope dish. 
After the new MAGIC-I camera was installed, counterweights on the back side of the structure had to be modified in order to compensate for the heavier weight of the new camera. 
Once works on the camera and the counterweights were finished, a new set of look up tables (LUTs) for the AMC were produced to achieve minimal optical PSF at every zenith angle (no dependence on azimuth) pointed by the telescope. 
The LUTs were produced by pointing the telescopes to stars at different zenith angles and minimizing the optical PSF (calculated from the reflected image of the star formed on a dedicated movable target positioned on the camera plane) by moving the actuators of the mirror panels.  
Images of stars are taken on night by night basis by a special high sensitivity CCD camera 
(SBIG$^{\textregistered}$\footnote{{\bf www.sbig.com}}) located in the center of the dish.  
A typical on-axis image defining the optical PSF  for both telescopes is shown in Fig.~\ref{fig:PSF}, where the 39\% light containment radius is 1.86' (1.80') and 95\% containment radius is 7.46' (6.51') for the MAGIC-I (MAGIC-II) telescope, respectively.   
With an increasing angle to the optical axis the PSF worsens following a second order polynomial function (see \citet{garczarczyk:2006:thesis}, figure~4.17). 
Note that the MAGIC camera pixel size has a dimension of 30\,mm flat-to-flat of the hexagonal entrance window of the Winston cone corresponding to a field of view of 6'.  
The stability of the PSF and the absolute reflectivity of the mirrors {\bf are subjects} of a forthcoming publication.

\begin{figure}[htb]
\begin{center}
\includegraphics[width=\linewidth]{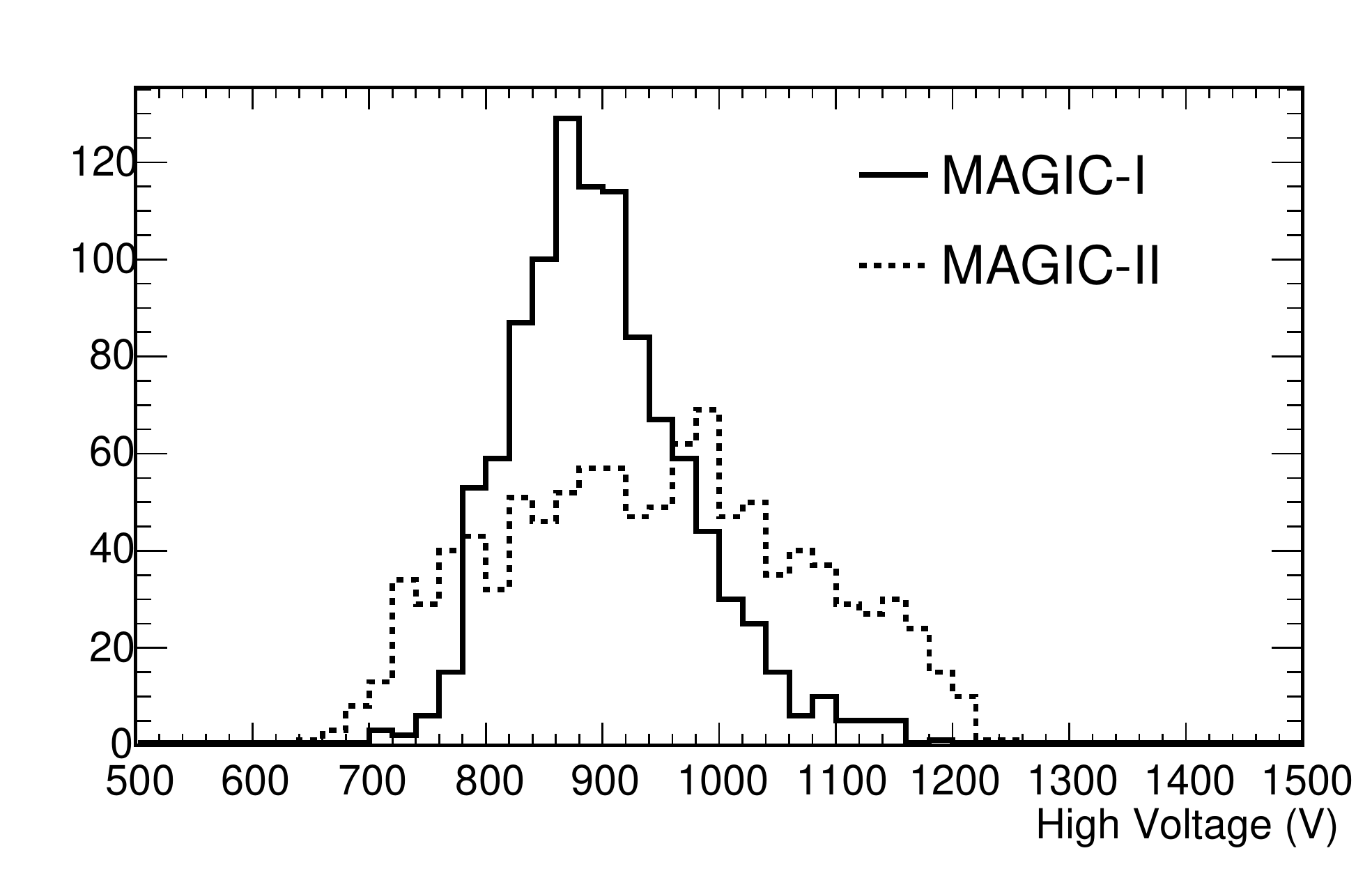}
\caption{
Distribution of the high voltages (HVs) applied to PMTs in MAGIC-I 
and MAGIC-II cameras after the charge flatfielding procedure.
See text for details.
The highest voltage that can be applied to the MAGIC PMTs is 1250\,V.}
\label{fig:HVs}
\end{center}
\end{figure}


\subsection{Flatfielding of the PMT gains}
\label{hvflatfielding}
\begin{figure}[htb]
\begin{center}
\includegraphics[width=1.0\linewidth]{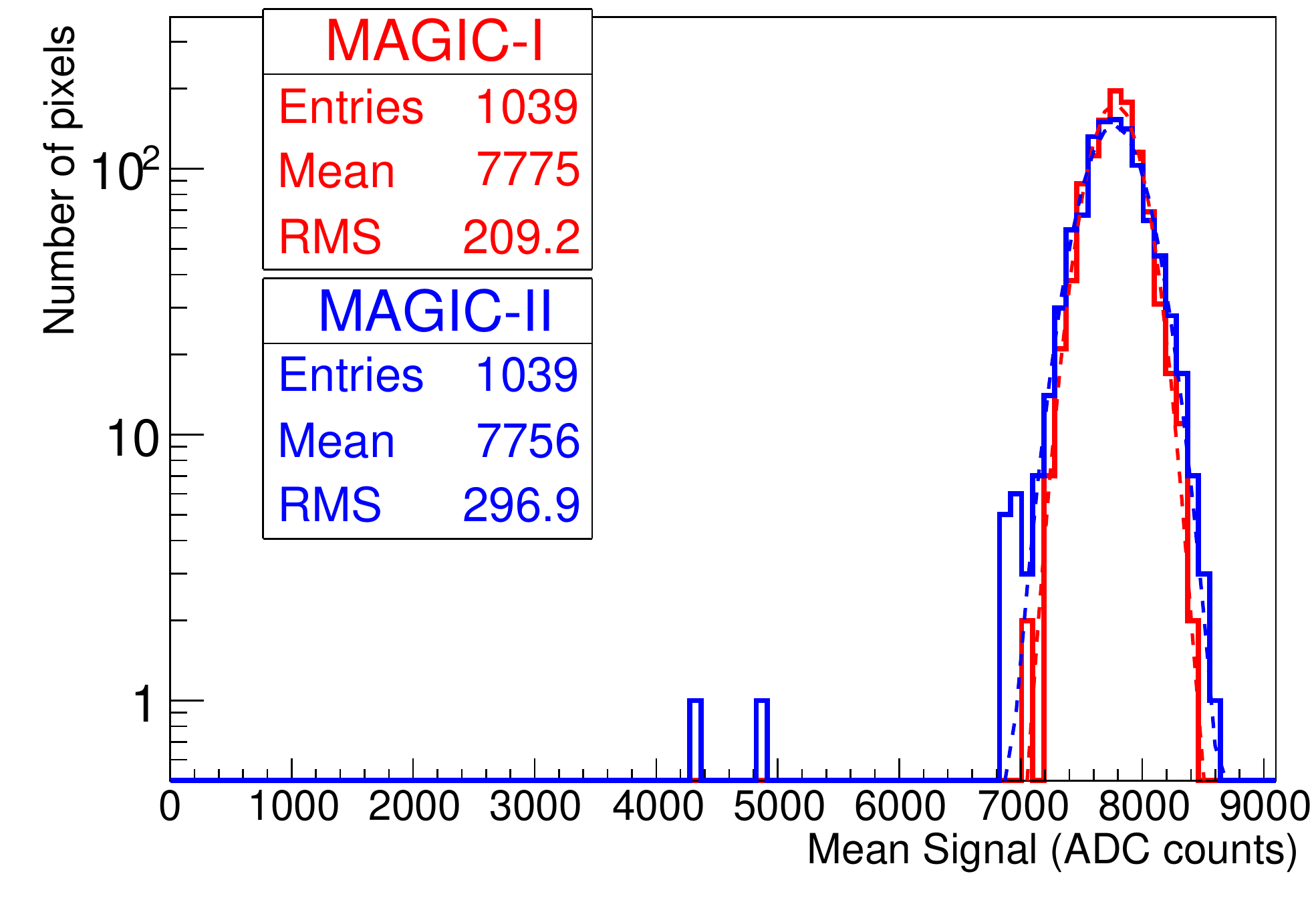}
\caption{
Charge distribution of the calibration pulses in the MAGIC-I (red) and MAGIC-II (blue) cameras after HV flatfielding.
As shown in the statistics, the mean values are very similar for the two telescopes whereas the spread (RMS) is 2\--4\%.
Data from 10 October, 2012.}
\label{fig:charge}
\end{center}
\end{figure}

Each PMT has a different gain at a fixed HV. 
The spread of the gains is unavoidable during the manufacturing process. 
We measured such spread in the PMTs for MAGIC~I and found that it is about 30-50\% (RMS),
depending on the production line.
The signal propagation chain introduces further differences in the gain: the optical links as well as the PIN diodes of the receivers mainly contribute to them. 
For the purpose of easier calibration of the signals and consistent saturation effects, the HVs applied to PMTs are adjusted such that the resulting signal from calibration pulses (equal photon number at the entrance of the PMTs) is equal in readout counts in all pixels when extracted after the digitization process.  
The adjustment of the HVs leads to differences in the transit time of the electrons in the PMTs. 
This is taken into account by automatically adjusting the delays of the L0 trigger signals (see {\bf Section}~\ref{trig_adjustment}).
The resulting HV distribution for MAGIC-I and MAGIC-II cameras can be seen in Fig.~\ref{fig:HVs}. 
The distribution of the MAGIC-I camera is narrower. 
This is due to the fact that during the construction of the MAGIC-I camera the PMTs were divided into two categories, high and low gain ones (see Section~\ref{magic-camera}).
The high-gain {\bf signals} are attenuated in the PMT base, which reduces the spread of the resulting gain distribution and consequently the spread of the HV distribution. 
The quality of the HV flatfielding can be seen in Fig.~\ref{fig:charge}, and results very similar for the two telescopes.
During operation, HVs are set once per night and typically not changed during the night, except in case of particularly bright light conditions such as strong moon light\footnote{
{\bf We refer to moon phases between the first and the last quarter.}}. 
The RMS (see inlay of the figure) values are similar to the $\sigma$ of the Gaussian fit (dashed lines), and reach to 2-4\% of the corresponding mean value.
There are two pixels in MAGIC-II that could not be flatfielded well because the gain is too low even at the maximum HV. 
These two pixels can still be used in the analysis, but with a lower signal to noise ratio.

\subsection{Trigger adjustments and validation} 
\label{trig_adjustment}
One of the most relevant systematic uncertainties of the detector originates from the camera's inhomogeneous response to $\gamma$ rays, starting from how they are triggered.
The inhomogeneity of the recorded Cherenkov pulses can come from different gains in the electronic chain, different electronic noise levels or different levels of the night sky background light (presence of stars in the FoV).
While the recorded pulses can be calibrated and flatfielded on the analysis level, the trigger inhomogeneity cannot be easily recovered.
Therefore, a special attention is given to make sure all channels in the trigger are working well, the discriminator thresholds (DTs) are flatfielded and all multiplicity combinations in the L1 trigger are properly functioning.
During the commissioning there were two major tasks concerning the L1 trigger:
\emph{(a)} validation of all next neighbor multiplicities  and \emph{(b)} the L0 delays are adjusted to assure that the time distribution of the Cherenkov photons in the focal plane of the telescope is conserved at the L1 trigger level, 
which is achieved by means of reference light pulses from the calibration box.
Dedicated hardware and software have been developed to test all multiplicities in short time. The L1 trigger systems of both telescopes have been extensively tested, hardware mistakes identified and repaired.

\subsubsection{Evaluation of the L1 trigger} 


The L1 trigger was evaluated with HYDRA, a multithread C-program to test, adjust and monitor the L1 trigger.
The program is running as a part of the MAGIC Integrated Readout (MIR) software, which is the slow control program to steer and monitor individual readout, trigger and calibration system components of the MAGIC telescopes \citep{tescaro:2009a}.
There are many trigger pixel combinations to test: 1653 for 2NN, 988 for 3NN, 1311 for 4NN
and 2280 for 5NN.
To test the L1 macrocell multiplicities, signals in all trigger channels are injected.
This can be done during the day thanks to the pulse injection system of the camera (see Sec.~\ref{magic-camera}).
The DTs are set below the injected signal for a particular pixel combination in every macrocell,
the others are set high enough to ensure that they will not trigger. 
The rate of the macrocell is monitored to identify inoperative combinations.
The algorithm checks all possible combinations sequentially but it runs in parallel for all 19 macrocells.
To go from one combination to the next one, the DTs must be changed, which takes about 10\,ms per pixel.
The trigger rate of the macrocells is read every 10\,ms, which makes the scan fast.
The procedure to test all possible trigger combination takes about 15\,min allowing for regular monitoring of the trigger performance. 
During the commissioning of the upgraded system, about 20 channels in each telescope were found not working in the L1 trigger (mostly due to a bad soldering and faulty components), which then were repaired.

\subsubsection{L0 delays and L0 width adjustment} 
\label{L0adjustement}
The arrival times of the signals at the L1 logic as well as the widths of the L0 signals had to be adjusted. The precision of the delay and width chips is 10\,ps.
There is a trade-off between the L1 trigger gate (that depends on the widths of the L0 signals) and the accuracy of the arrival times adjustment.  
With no delay adjustment the time spread would have an RMS of 3--4\,ns with some outliers up to 10ns. 
The spread of arrival times is due to differences in transit times of the electrons in PMTs (mainly because of different HV applied) and to differences in signal travel time through
optical fibers, as well as slightly different response time of electronic components.  
Finally, {\bf in this case} not related to the hardware, one also needs to allow for some
2\,ns differences in arrival time between individual channels due to the physics of the showers\footnote{The time gradient can be up to 2\,ns between
neighboring pixels for high-energy showers with a large impact distance from
the telescope. This is a pure physics effect that cannot be tuned/minimize as the other time differences we discussed in this section.}.
Without delay adjustment of individual channels the L1 trigger gate would, therefore, be at least 15\,ns to secure maximal efficiency of the L1 coincidence trigger to $\gamma$~rays. 
A larger gate corresponds to a higher chance to receive an accidental trigger, and the accidental trigger rate is a factor limiting the energy threshold 
since discriminator thresholds have to be increased to keep their rate under control.  
The goal was, therefore, to keep the gate as low as possible by adjusting the arrival times between the channels. 
In the following we describe the approach we used.

\begin{figure}[htb]
\begin{center}
\includegraphics[width=\linewidth]{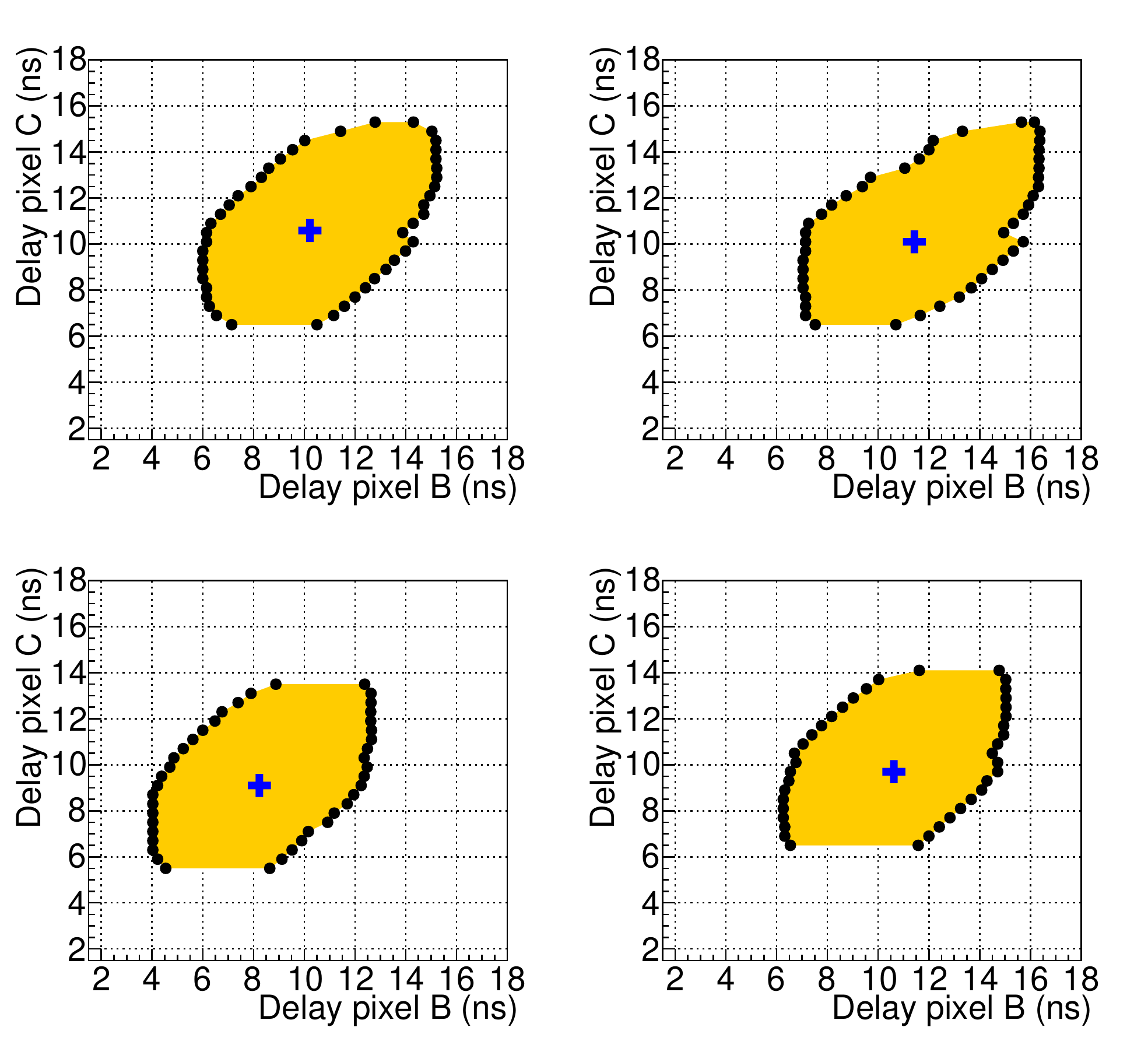}
\caption{
Example of the adjustment of the L0 delays between neighboring pixels of the 3NN logic.
Shown are 4 macrocells with an example of 3NN combination between pixels pA, pB and pC.
The result of the scan is shown by the blue cross. See text for details.}
\label{fig:3nnL0s}
\end{center}
\end{figure}

\begin{figure}[htb]
\begin{center}
\includegraphics[width=\linewidth]{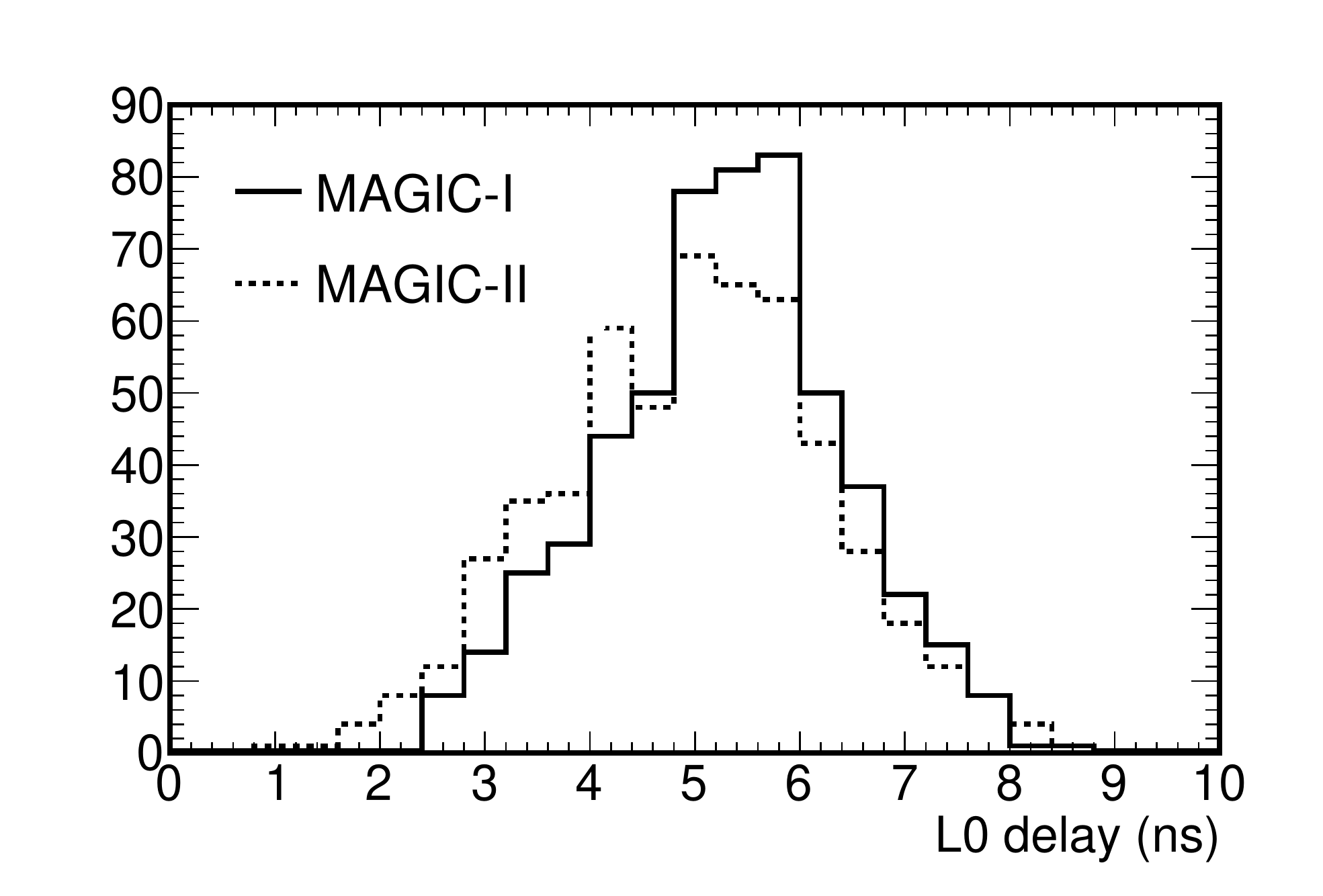}
\caption{
Distribution of the L0 trigger delays in MAGIC-I 
and MAGIC-II after applying the HYDRA optimization procedure.} 
\label{fig:L0s}
\end{center}
\end{figure}

In HYDRA, several algorithms were implemented to adjust the arrival times automatically. 
Since the transit time in the PMTs has a relevant contribution, the procedure must be performed with the flatfielded HVs and open camera, using calibration pulses (since they arrive simultaneously at the camera plane, see Section~\ref{MAGICCalibSystem}). 
The following algorithm was chosen to be the standard one: the adjustment is done in 3NN logic and a fixed L0 pulse width.  
In every macrocell, the central pixel of the macrocell is considered to be the reference channel.
A 2D scan in delay times is performed with the two neighboring pixels (in a valid 3NN combination) and the delays are chosen to maximize the resulting L1 gate.  
An example of such 2D scans for 4 different macrocells with a particular 3NN combination between pixels pA, pB and pC is shown in Fig.~\ref{fig:3nnL0s}. 
The delay of pixel pA is kept constant whereas a scan in delays of pixels pB and pC is performed.
The axes of the plots indicate pixel delays in ns.  
The yellow area marks the delay combinations that result in a valid L1 trigger.  
The blue cross in the center of the area corresponds to the chosen delays as the result of the scan, 
and its position is defined as the crossing point between the maximal intervals in both directions (within some tolerances). 
If there are several possibilities, the mean values are taken.
The resulting L1 gates, the maximum width/height of the yellow area, are in {\bf the} order of $(7\pm1)$\,ns (Fig.~\ref{fig:3nnL0s}). 
The procedure continues successively over the 3NN combinations of the macrocell in a spiral by keeping already adjusted delays fixed. 
At the end, a cross-calibration procedure between the macrocells is applied using the border channels that belong to more than one macrocell. 
We also apply an overall offset to the resulting L0 delays to align them at around 5\,ns to minimize the trigger latency. 
The overall precision of the adjustment is $\pm1$\,ns (RMS). 
The procedure has been tested with different L0 signal widths finding that 5.5\,ns FWHM
is the shortest L0 signal that give robust and reproducible results with a high trigger homogeneity. 
Although the setting of the delays can not be done truly in parallel (since the communication bus is serial), the operation is very fast (1\,$\mu$s), so that in practice parallelizing the adjustment for the 19 macrocell gives a factor $\sim$19 gain in execution time.
The HYDRA-based procedure to adjust L0 delays (needed every time HVs are changed) takes about 15\,min, which is a substantial improvement compared to the former manual procedure that required several observing nights to finish.
The resulting L0 delays are shown in Fig.~\ref{fig:L0s}.

\subsubsection{Discriminator threshold (DT) calibration}
\begin{figure}[htb]
\begin{center}
\includegraphics[width=\linewidth]{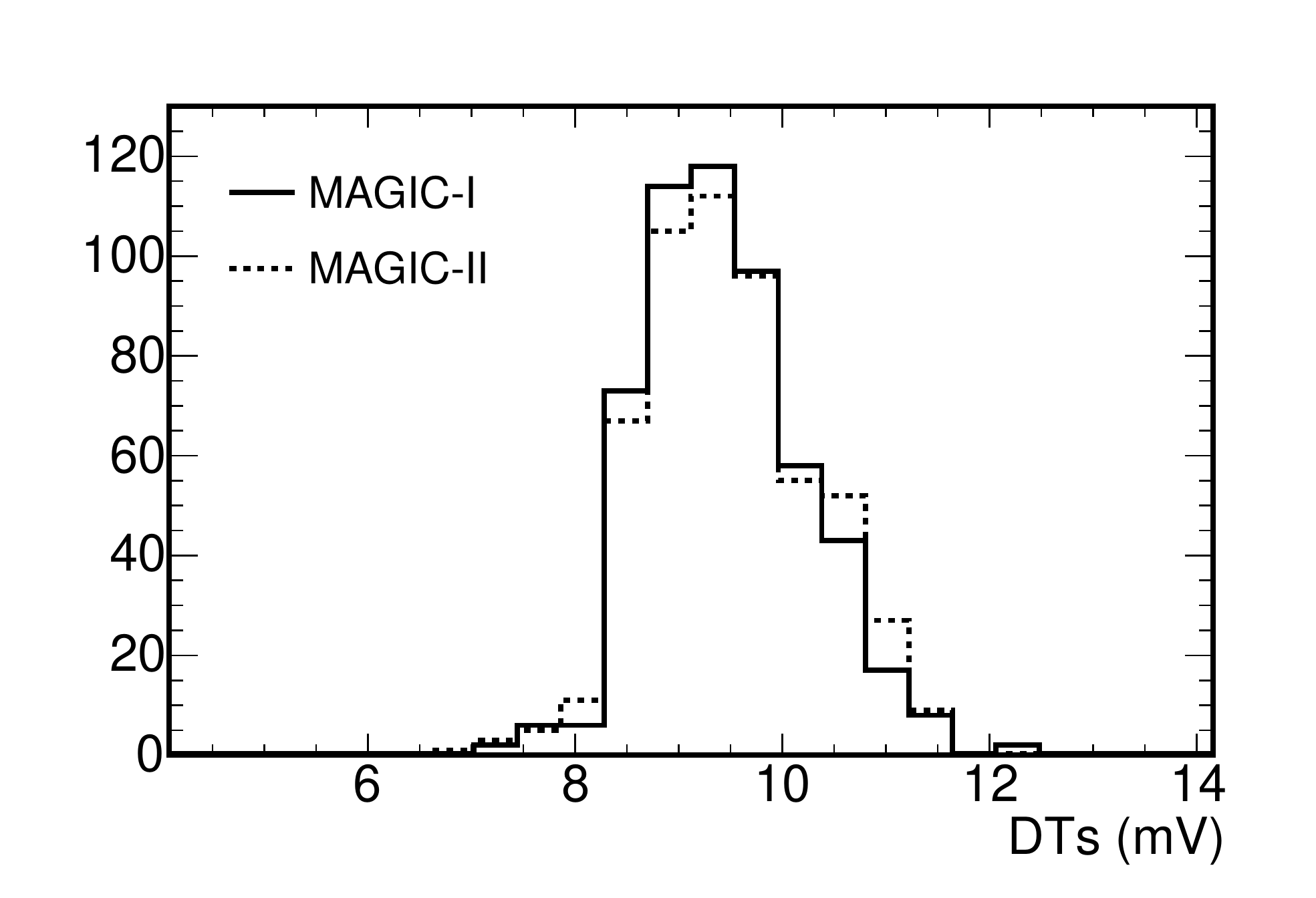}
\caption{
Distribution of the L0 discriminator thresholds (DTs) applied in MAGIC-I 
and MAGIC-II receiver boards after the charge flatfielding procedure
{\bf in order to achieve a charge threshold of 4.25\,phe.}}
\label{fig:DTs}
\end{center}
\end{figure}

The goal of the DT calibration is to adjust the DTs of the trigger channels (L0 trigger)
such that the sensitivity of the channels is flat in terms of photon density of Cherenkov
photons. This is achieved by means of a rate scan over the range of DTs for
each trigger channel when firing 
calibration light pulses with a given photon
density (e.g., equivalent to a mean of 100 photoelectrons (phe) per pixel in the PMTs of the camera).
Then, for each trigger channel the required DT is determined such that 50\% 
of the calibration pulses fulfill the trigger condition. 
These are the DTs corresponding to a 100 phe level. 
We then scale the DTs linearly to obtain individual pixel thresholds for a desired phe level. 
The extragalactic dark sky\footnote{
We define ``extragalactic dark sky settings" for an observation outside the Galactic plane during a dark night. 
DTs values can be scaled up in case of a FoV inside the Galactic plane or in case of brighter light conditions (e.g. with the moon in the sky).} 
DTs are set to be at a level of 4.25\,phe. 
For Galactic sources, 15\% higher DTs are used.
Before the MAGIC upgrade, the extragalactic dark sky settings were 4.3\,phe in MAGIC-I and 5.0\,phe in MAGIC-II,
respectively \citep{aleksic:2012:magic:stereo:performance}.
The change in the operation DTs shows that MAGIC-II was improved for operation at a lower threshold and MAGIC-I
threshold was maintained constant despite the larger trigger area. 
The distribution of the post-upgrade DTs for the two telescopes can be seen in Fig.~\ref{fig:DTs}. 
The spread in the DTs for a homogeneous light is about $\sim$15\% RMS and is higher than the spread of the charges (2-4\%, see Fig.~\ref{fig:charge}). 
The reason is that there are some small differences between the analog (readout) and digital (trigger) signal branches: 
the signal shapes are not identical,
and the DT is applied to the amplitude, whereas the charge is extracted {\bf from} the integrated signal.
\\

\subsubsection{Individual pixel rate control (IPRC)}
\label{sec:IPRC}

During operation, the individual pixel rates (IPRs) are dominated by the night sky background. 
Bright stars inside the FoV illuminate small areas of the camera and increase the IPR of the affected pixels. 
The flatfielded DTs may result in very different IPRs during the operation since \emph{(a)} the response of the PMTs is different for the calibration pulses for which the DTs are calibrated (fixed wavelength 355 nm) and the night sky background (roughly a power law spectrum growing to red wavelengths)
and \emph{(b)} the rates depend on the sky region the pixel is exposed to (e.g.\ it may contain stars, which would increase NSB fluctuations and, therefore, the IPR). 
During the commissioning, several algorithms and limits were tested and the following procedure established: As long as the IPRs are below 1.2\,MHz
(most of them being in the range of 300--600\,kHz), no action is taken. 
For an IPR outside of this limit (due to {\bf the} presence of stars), an IPR control software takes care of increasing the DT for the affected pixel in order not to spoil the resulting L1 telescope rate 
(the channels are typically still suitable for image analysis though). 
The DT is increased until the pixel rate falls within the limits of 200\,kHz -- 1.2\,MHz.
In case of really bright stars in the pixel FoV (typically with B-magnitude {\bf higher} than {\bf mag} 3 causing a DC level higher than 47\,$\mu$A) the applied HV is automatically reduced to protect the PMT from fast aging.
Once the star is out of the FoV of the affected pixel, its IPR will be low because of the previously increased DT and once the IPR is below 100\,kHz the IPR control (IPRC) software resets the DT to the original value. 
This procedure ensures a flexibility for different NSB levels and takes care of the stars in the FoV while keeping the energy threshold low and most of the DTs flatfielded. It is important to keep the DTs flatfielded to ensure a good matching between the data and Monte Carlo simulations, where it is assumed that the DTs are identical for all the pixels.
The number of pixels affected by the procedure depends on the sky field. 
If it does not contain many stars with magnitude {\bf higher} than {\bf mag} 8 there are typically have 1-10 pixels per camera, for which DTs are modified. 
For sky fields with many stars we can easily have up 100-150 affected pixels per camera.

\subsubsection{Adjusting the operating point of the trigger}
\label{l3triggeradjustment}

\begin{figure}[t!]
\begin{center}
\includegraphics[width=1.0\linewidth]{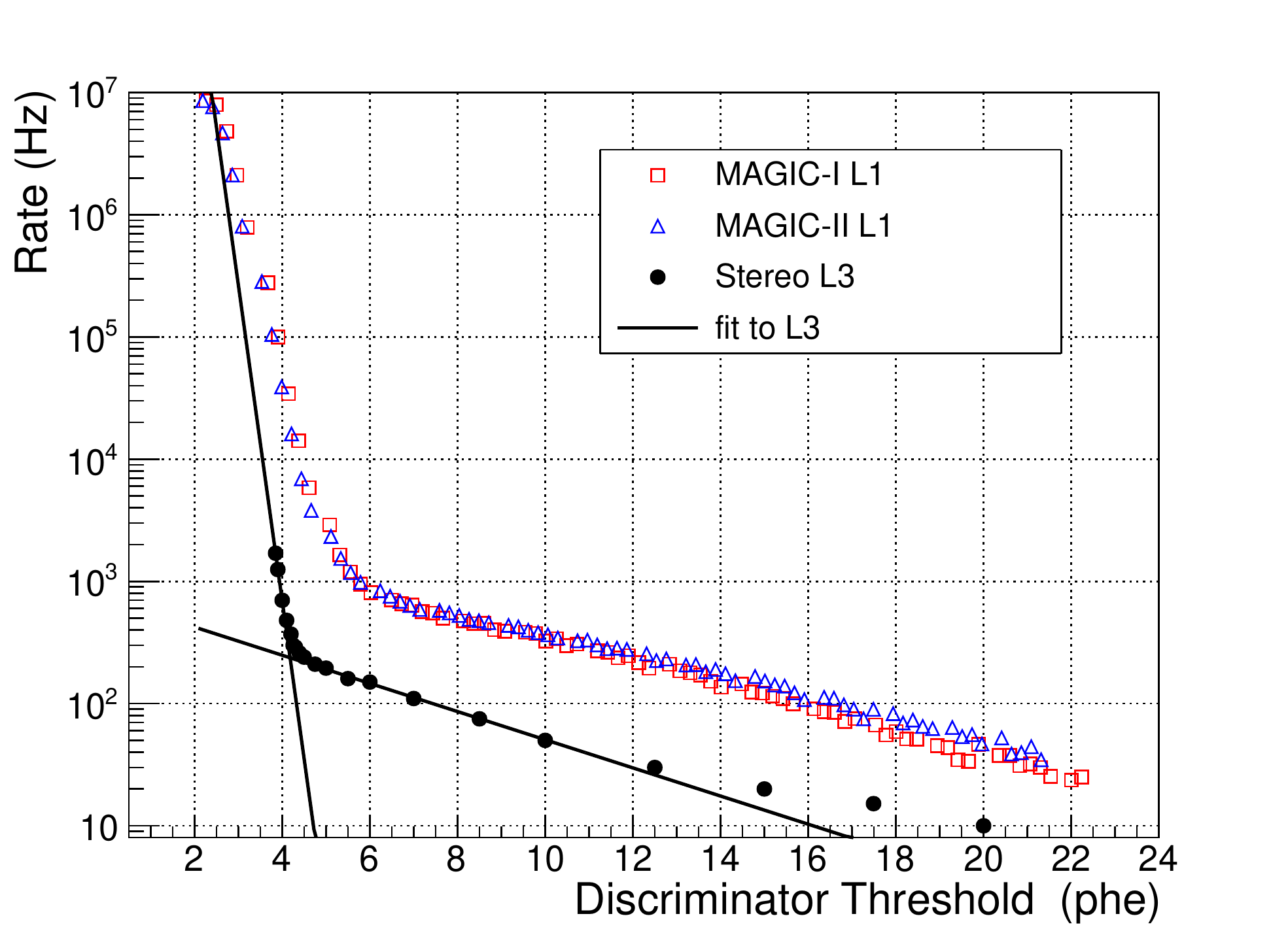}
\caption{
Rate scans taken by changing discriminator thresholds (DTs) to optimize the operating point of the MAGIC telescopes.
Red (blue) points are L1 3NN rate scans taken with MAGIC-I (MAGIC-II) telescopes and the lines are analytical fits to them.
The black points correspond to the measured stereoscopic rate of the system,
and the black lines are fitted functions. The operating point has been chosen to be 4.25\,phe per channel.}
\label{fig:ratescan}
\end{center}
\end{figure}

Rate scans have been performed at clear nights at low zenith angles to determine the trigger rate as a function of the DTs in phe. 
Mono (L1) trigger rate scans as well as stereoscopic (L3) trigger rates scans have been performed for several nights and the performance has been found to be stable.
An example of the rate scans is shown in Fig.~\ref{fig:ratescan}. 
One can see the steep slope of the rate at low DTs, where the rate is dominated by the chance coincidence due to NSB and afterpulsing of PMTs. 
At higher DTs, the trigger rate is dominated by the rate of the cosmic ray showers and penetrating muons. 
One can see that the coincidence trigger (L3 trigger) strongly suppresses the chance coincidence triggers
and the triggers due to local muons. However, the L3 trigger rate is also lower (the factor is energy dependent) 
{\bf since, for any fixed individual telescope rate, the stereo collection area is smaller than the mono one because 
it is an intersection of the mono trigger areas of the two telescopes.}
For standard operation, the L1 3NN trigger logic and the hardware stereo (readout of the camera only for stereo triggers) is used.
The operating point for the L0 trigger has been chosen to be 4.25\,phe, resulting in a stereo rate of around 280\,Hz, of which about 40\,Hz are accidental triggers.
\\

When pointing to new observation targets, it takes several seconds for the IPRC to adjust the DTs to the NSB light in the FoV.
In the commissioning of the system we noted that, during these short periods, the resulting stereo trigger can be very high (5-100\,kHz), being dominated by accidental triggers.  
To avoid possible saturation of the data acquisition with very high trigger rates, which may lead to data corruptions and interruption of data taking procedure, a trigger limiter was installed inside the prescaler\footnote{The prescaler board is used in the MAGIC telescope to select the triggers coming from the individual trigger sources before issuing a trigger signal to the readout system.} of each telescope.  
The trigger limiter evaluates the overall trigger rate of a telescope every 10\,ms and blocks triggers to the readout for the following 10\,ms in case the rate is above a programmable limit. The trigger limiter is configured to block trigger rates above 1000\,Hz.

\subsection{DAQ performance}

The total acquisition rate that the DAQ has to guarantee is the sum of the cosmic trigger rate provided by the L3 trigger ($\sim$280~Hz, see section \ref{l3triggeradjustment}) plus the contribution of the diagnostic calibration and pedestal triggers (25+25\,Hz) that are issued interleaved to the physical triggers.
During the commissioning of the system we adjusted various parameters of the DAQ program such as the number of cores used, loads between different threads, DAQ buffer length etc.
The total sustainable acquisition of the DAQ system is currently $\sim$800~Hz (CPU limited, the writing speed being $\sim$1.1~kHz), more than double than the actual data taking rate.

The CPU overhead is particularly demanding because besides the event building, the data integrity check and the data storage, the DAQ performs two further
actions: the DRS4 raw data correction and the extraction of the online data check values. 
The raw data correction (see Section~\ref{pre-processing}) is a particularly demanding task because it involves the manipulation of every single digitized data sample from the readout, that has to be pedestal-subtracted using the specific single DRS4 capacitor average pedestal value \citep{sitarek:2013a}.

As discussed in Section~\ref{sec:receiverboards}, the receiver boards of the upgraded system allow a very precise control of the trigger rate even when the light conditions change
(IPRC of the L0 discriminators).
Nevertheless, besides providing this sustainable average acquisition rate, the DAQ has to guarantee a certain tolerance to sudden increase of the acquisition rate\footnote{Temporary increases of the trigger rate are typically determined by uncontrolled factors like car-flashes from the astronomers of the observatory, but might be potentially also due to particularly high 
$\gamma$-ray fluxes from extremely bright flares.}.  
This capability is guaranteed---without event loss---by the DAQ program design, which relies on a volatile memory ring-buffer bridging the events to the disk. 
The event buffer is 10,000~events deep, or 10~s in time at a trigger rate of 1~kHz.

While running, the DAQ also feeds the online analysis program dubbed MOLA \citep[see][and Section~\ref{mola}]{tescaro:2013a}.

\section{The online analysis client} 
\label{mola}

A real time data analysis is an important part of the success of an IACT experiment.
Most of the extragalactic and several Galactic very-high-energy sources are variable, some of them on time scales down to hours and minutes. 
A real time analysis of the data taken can provide essential time critical internal triggers to extend observation of flaring sources and alert other multiwavelength partners.

The upgraded system allowed to develop a novel program to fulfill the task of analyzing the data as they are being taken (contrary to the on-site analysis mentioned in Section~\ref{computing} that starts when the night is over), 
and provide online information to the observers of the measured $\gamma$-ray flux and its time evolution: MAGIC Online Analysis (MOLA). 

MOLA is a multithreaded C++ program that runs simultaneously with the data acquisition software and acts as a receiving client of the event informations computed at the very moment the events are acquired by each telescope. 
In fact, as mentioned in the previous section, the DAQ software of MAGIC-I and MAGIC-II computes independently the signal and arrival time of each pixel of the telescope cameras.
In this way the calculation of the image parameters and the latest steps of the data analysis are outsourced to a separate program on an independent computer.

The multithread program structure consists of three threads: two \emph{reading} threads and one \emph{analyzing} thread. 
The two reading threads are appointed to receive the data stream via TCP/IP from the two DAQs asynchronously and perform the non-stereo analysis steps. 
The main analysis thread is instead appointed to match the events from the two streams and perform the stereoscopic reconstruction (see below).

Two independent TCP/IP streams are activated once the program starts, and each time the observation of a source is finished the current results are stored and the analysis reset.
Each event stream contains the pixel signals (integrated charge) and the signal arrival time.
The tasks of each reading thread are:
\begin{itemize}
\item Receive from the DAQ program, decode and temporarily store the relevant information from the event stream (event tags, charge per pixel and relative arrival time).
\item Calibrate (flat-field) the gain and identify the dead pixels using interleaved calibration events.
\item Check pedestal events to identify and interpolate signals from intrinsically noisy or dead pixels
(e.g.\ pixels with hardware problems, typically less than 5 per camera).
\item Perform image cleaning to select pixels with a significant Cherenkov signal.
\item Calculate image parameters, using standard MAGIC analysis software data structures 
\citep[see][]{zanin:2013:magic:mars}.
\item Estimate shower direction from a set of relevant parameters (image
shape, orientation and time gradient along the major axis) by 
{\bf means of the Random Forest classification as described in} \citet{albert:2008:magic:RF}. 
\end{itemize}

Single telescope events have to be combined to form stereo events in order to exploit the full potential of the stereo imaging technique. 

The tasks which have to be accomplished in order to obtain high level analysis results are performed by the stereo analysis thread, and can be summarized as follows:
\begin{itemize}
\item Identify matching stereo events by means of the unique L3 trigger number.
\item Calculate the event direction through a weighted average of the estimates from the two individual images.
\item Calculate shower core impact point and impact parameters.
\item Apply the background suppression by means of the \emph{hadronness} gamma/hadron likelihood parameter 
\citep{zanin:2013:magic:mars}.
\item Apply cuts and compute the signal excess plot with respect to the candidate source position.
\item Produce sky-maps with $\gamma$-ray candidate events.
\item Produce light curves (time evolution) of the measured $\gamma$-ray flux during the current observations. 
\end{itemize}

Results are produced for two energy ranges: Low Energy (LE) and High Energy (HE) depending of the size of the event image in phe:
The HE sample includes all events with at least 125\,phe in each of the two telescopes; 
the rest of the events with at least 40\,phe in each telescope constitutes the LE sample. 
For Crab~Nebula low zenith angles ($<30^\circ$) observations the median energies of these two samples are $\sim$110 and $\sim$350\,GeV, respectively.
In the HE range, the sensitivity of the MOLA analysis at zenith angles below 25$^\circ$ zenith angle has been 
estimated to be 1.0\% of the Crab~Nebula flux in 50\,h observation time, which is equivalent to 10\% of the Crab~Nebula flux in 30 minutes.

MOLA provides to the telescope operators high-level information about the currently observed astrophysical source such as signal excess plots and sky-maps with $\gamma$-ray candidate events, together with diagnostic information related to the signal calibration and the image parameters calculation.
MOLA is commissioned to perform without data loss up to a rate of 600~Hz. For higher data rates (not expected with the current setup), some events will be lost for the online analysis but the program will continue running with a reduced performance.


\section{Conclusions}

A major upgrade of the MAGIC telescopes took place in the years 2011--2012. 
The major items were the installation of the new camera for MAGIC-I, the new trigger in the MAGIC-I telescope, the upgrade of the readout system to DRS4 and programmable receiver boards in both telescopes. 
The commissioning of the upgraded system successfully finished in October 2012, and the telescopes restarted regular operation. 
The main goals of the upgrade were an improvement of the sensitivity at low energies, unification of the hardware used, and reduction of down time due to technical problems. 
These goals have been successfully achieved, e.g.\ the down time due to technical problems was $<$\,10\% of the observation time 
in the first two years after the upgrade was finished,
with only less {\bf than} 2\% observation time loss due to troubles in the upgraded subsystems.
This is more than a factor of two better than in the years before the upgrade and at the level required 
for the new generation Cherenkov telescopes of the CTA observatory \citep{cta:concept:2013a}.
The expectations concerning the sensitivity of MAGIC were conservative since they have been surpassed.
A comprehensive comparison can be found in the Part~II of this paper, \citet{aleksic:2014:magic:upgradePartII}.

\section{Acknowledgements}
We would like to thank the Instituto de Astrof\'{\i}sica de Canarias for the
excellent working conditions at the Observatorio del Roque de los Muchachos in
La Palma. The financial support of the German BMBF and MPG, the Italian INFN
and INAF,  the Swiss National Fund SNF, the ERDF under the Spanish MINECO, and
the Japanese JSPS and MEXT is gratefully acknowledged. This work was also
supported by the Centro de Excelencia Severo Ochoa SEV-2012-0234, CPAN
CSD2007-00042, and MultiDark CSD2009-00064 projects of the Spanish
Consolider-Ingenio 2010 programme, by grant 268740 of the Academy of Finland,
by the Croatian Science Foundation (HrZZ) Project 09/176 and the University of
Rijeka Project 13.12.1.3.02, by the DFG Collaborative Research Centers
SFB823/C4 and SFB876/C3, and by the Polish MNiSzW grant
745/N-HESS-MAGIC/2010/0.  We thank the two anonymous referees for thorough
reading and helpful comments on the manuscript.


\bibliographystyle{elsarticle-harv}
\bibliography{ms}


\end{document}